\documentclass[prd,letterpaper,tightenlines,nofootinbib,preprint,floatfix,showpacs]{revtex4}
\newcommand{\Eq}[1]{Eq.~(\ref{#1})}
\renewcommand{\r}[1]{(\ref{#1})}
\newcommand{\z}[1]{\left({#1}\right)}
\newcommand{\kz}[1]{\left\{{#1}\right\}}
\newcommand{\sz}[1]{\left[{#1}\right]}
\newcommand{\m}[1]{\mathrm{#1}}
\renewcommand{\v}[1]{\mathbf{#1}}
\newcommand{\rec}[1]{\frac{1}{#1}}

\usepackage{graphicx}
\usepackage{epsfig}
\usepackage{color}

\def\bea{\begin{eqnarray}}
\def\eea{\end{eqnarray}}
\def\be{\begin{equation}}
\def\ee{\end{equation}}

\begin{document} 
\title{Observables and initial conditions for rotating and expanding
fireballs with spheroidal symmetry
\footnote{Dedicated to L. P. Csernai on the occasion of his 65th birthday}
}
\author{T. Cs{\"o}rg{\H o}$^{1,2}$, M.I. Nagy$^3$  and I. F. Barna$^1$}
\affiliation{$^1$MTA WIGNER FK, H-1525 Budapest 114, POB 49, Hungary}
\affiliation{$^2$KRF, H-3200 Gy{\"o}ngy{\"o}s, M\'atrai \'ut 36, Hungary}
\affiliation{$^3$ELTE, H-1118 Budapest XI, P\'azm\'any P. 1/A,  Hungary\\}

\begin{abstract}
Utilizing a recently found class of exact, analytic rotating solutions of non-relativistic fireball hydrodynamics, we calculate analytically the single-particle spectra, the elliptic flows and two-particle Bose-Einstein correlation functions for rotating and expanding fireballs with spheroidal symmetry.   We demonstrate, that rotation generates final state momentum anisotropies even for a spatially symmetric, spherical initial geometry of the fireball. The mass dependence of the effective temperatures, as well as the HBT radius parameters and the elliptic flow are shown to be sensitive not only to radial flow effects but also to the magnitude of the initial angular momentum.
\end{abstract}

\date{\today}


\maketitle

\section{Introduction} 

In non-central heavy ion collisions, the impact parameter is non-vanishing, and
the initial conditions include a non-vanishing, but frequently neglected
initial angular momentum of the nucleons that participate in the inelastic
collisions.  This non-zero initial angular momentum of the participant zone is
a conserved quantity that survives the initial, non-equilibrium stage of the
heavy ion collision, that results in thermalization.  The fireballs created in
non-central high energy heavy ion collisions will thus not only expand but
rotate as well. The vast majority of numerical and analytic hydrodynamical
calculations performed so far completely neglected the effect of the initial
angular momentum on the final state observables. The effects of rotation,
however, may be rather significant and, rather recently, the imprints of rotation
on the observables started to draw significant theoretical attention.

As far as we know, the initial angular momentum of non-central heavy ion
collisions was first taken into account in ref.~\cite{Csernai:2011gg}, where a
sign change of the directed flow $v_1$ was predicted and related to the change
of the relative importances of the angular momentum driven expansion to the
pressure driven radial expansion.  The rotation effect was also related to a
rapidity dependent oscillating pattern of the elliptic flows when the
fluctuations of the initial center-of-mass rapidity are corrected for.  These
results were highlighted as an important feature of the description of the
nearly perfect fluids formed in high energy heavy ion reactions at RHIC and LHC
energies using a numerical hydrodynamical model, that is based on the particle
in cell method~\cite{Cifarelli:2012zz}.  This hydrodynamical model is a
three-dimensional, finite calculation, and it was tested against numerical
viscosity and entropy production artefacts ~\cite{Csernai:2011gg}.  The
finiteness of the fireball is an important requirement, 
when rotation effects are considered, given that models that assume 
longitudinal boost invariance and flat rapidity
dependence have infinite moment of interia in the beam direction, hence they
cannot, by definition, take into account the explosion and simultaneous
rotation of the fireball in the impact parameter plane.  Such a rotation, by
definition,  would require a time dependent and finite moment of inertia. 

The differential two-particle Bose-Einstein or Hanbury Brown -- Twiss 
correlation function (DHBT) was proposed as an effective
tool to measure the angular momentum of non-central heavy ion collisions
~\cite{Csernai:2013vda}. The results detailed in ref.~\cite{Csernai:2013uda}
indicated that the DHBT method can indeed be used to detect the angular
momentum and the rotation of the fireball, but the effects of irregular initial
shapes, density fluctuations, irregular radial flows require extended analysis
to disentangle these effects from one another.  The differential correlation
function was numerically found to dependent on the shape, the  temperature, the
radial velocity and the angular velocity as well as on the detector position,
as it was demonstrated in refs.~\cite{Velle:2015xha}, and the numerical value
of the DHBT correlation function for realistic angular velocities and radial
flows was found to be rather small, of the order of 2-3 \% .  Increased initial
angular momentum of the rotating and expanding fireball was shown to result in
an effective decrease of the observable correlation radii in
ref.~\cite{Velle:2015dpa}. 

These numerical evaluations of observable signals of a non-vanishing initial
angular momentum were partially based on a recently discovered family of exact
solutions of rotating and expanding, non-relativistic fireball hydrodynamics,
detailed in ref.~\cite{Csorgo:2013ksa}.  This solution with non-vanishing
initial angular momentum can be considered as the rotating and spheroidally
symmetric generalizations of the radially expanding, finite, Gaussian exact
solution of fireball hydrodynamics found already in
1998~\cite{Csizmadia:1998ef}.  That solution was shown to be a simultaneous
solution of the equations of non-relativistic fireball hydrodynamics with
spheroidal symmetry, and at the same time, also a solution of the
non-relativistic form of the collisionless Boltzmann equation. The first
relativistic solution of fireball hydrodynamics for rotating fluids was found
by generalizing this method of ref.~\cite{Csizmadia:1998ef} to relativistic
kinematics, i.e. looking for those families of solutions of relativistic
fireball hydrodynamics that also simultaneously solve the collisionless
relativistic Boltzmann equation~\cite{Nagy:2009eq}. It is interesting to note
that several families of exact, rotating hydrodynamical solutions were found in
this class, for example rotating Hubble flows and rotating but asymptotically
non-Hubble flows as well. Hatta, Noronha and Xie rediscovered these solutions
independently and generalized them also for axially symmetric, expanding
fireballs with non-vanishing viscosity ~\cite{Hatta:2014gqa}, carrying out a
systematic search using AdS/CFT correspondence techniques for similar solutions
as well~\cite{Hatta:2014gga}. 
The influence of a non-vanishing initial angular momentum was also investigated in
the holographic picture of Quark Gluon Plasma~\cite{McInnes:2014haa}, pointing
out that the estimates of quark chemical potential can be considerably improved
by taking the angular momentum conservation into account.  The quickly
developing field of exact and analytic solutions of rotating fireball
hydrodynamics was briefly reviewed in ref. ~\cite{deSouza:2015ena}, that also
discussed recent efforts to consistently derive and formulate the theory of
viscous relativistic hydrodynamics, as well as emphasized the conceptual
difficulties that relate to the application of the hydrodynamical method to
high energy heavy ion collisions, summarizing at the same time the enormous
successes of the hydrodynamical models at RHIC and LHC energies.

Due to the conservation of angular momentum, and assuming the validity of
certain kind of equipartition theorem, ref.  ~\cite{Becattini:2013vja}
predicted that the $\Lambda$ and $\overline{\Lambda}$ baryons emerge from
the non-central heavy ion collisions in a polarized manner. Using numerical
hydrodynamical calculations as well as the analytic approach of ref.~\cite{Csizmadia:1998ef}, the $\Lambda$ polarization was evaluated by taking into account
both the radial expansion and the rotation effect simultaneously ~\cite{Xie:2015xpa}
and an observable amount of $\Lambda$ polarization was predicted.
In addition, ref. ~\cite{Csernai:2015jsa} evaluated the vorticity from the
exact rotating solutions and pointed out, together with ref.
~\cite{Csernai:2014hva} the importance of the Kevin-Helmholz instability
in the initial stages of the fireball evolution.

Despite these theoretical efforts, as far as we know, the ongoing experimental
studies could not yet separate the angular momentum effects from radial flow
effects on spectra, elliptic flow and Bose-Einstein correlations, partly
because there was a lack of clear theoretical understanding how the
non-vanishing value of the initial angular momentum influences the final state
observables, that are typically measured and interpreted as variables sensitive
to radial flows.  Elliptic flows are particularly fashionable observables, that
are deeply connected to the fluid nature of the quark matter created in heavy
ion collisions.  They are frequently interpreted in terms of pressure gradients
and radial flows, that convert the initial spatial anisotropy to final state
momentum space anisotropy. 

In this manuscript, we consider the case of the recently found analytic, exact
solutions of non-relativistic fireball hydrodynamics~\cite{Csorgo:2013ksa},
that describe rotating expansions with spheroidal symmetry. In particular we
consider rotating expansions of a spheroid, where the spheroid under
investigation is such an ellipsoid, whose principal axes perpendicular to the
angular momentum are equal.  Our main goal is to clearly demonstrate the
influence of the initial conditions, in particular the non-vanishing value of
the initial angular momentum of the fireball, on the final state observables.
Utilizing this solution, we derive simple and straight-forward analytic
formulae that provide a possibility to experimentally test the effects of
rotation on fireball hydrodynamics.  We also investigate the dependence of the
observables on  the freeze-out temperature.  Although our treatment of the
rotation of the expanding fireball in heavy-ion collisions definitely
over-simplifies the physical situation of non-central heavy ion collisions, to
our knowledge, this is the first successful attempt to analytically determine
the effect of the rotation on the single particle spectra, elliptic flow and
HBT radii for finite, expanding fireballs. 

The structure of this manuscript is as follows. In section~\ref{s:solution} we
recapitulate the rotating and expanding solution of fireball hydrodynamics that
we use for the evaluation of the observables and present a generalization of
its first integrals of motion.  
In section~\ref{s:obs} we derive the analytic formulae that describe
the single particle spectra, elliptic and higher order flows as well as the
various azimuthally sensitive HBT radii for this family of rotating and
expanding exact solutions of fireball hydrodynamics. We illustrate the analytic
results also by numerical calculations of an exploding and rotating, initially
spherical fireball, and demonstrate how larger and larger initial angular
momentum may influence more and more the slope of the single particle spectra,
the particle mass and transverse momentum dependence of the elliptic flow and
the azimuthally sensitive HBT radii. Finally, we summarize and conclude.

For the sake of completeness, the manuscript is closed by two Appendices.
Appendix~\ref{s:app:H} details how we reduce the evaluation of the
observables from these rotating solutions of fireball hydrodynamics to
integration by quadratures, and gives the conditions of validity for these
derivations.  To advance the knowledge of possible hydrodynamical solutions
that might be useful for future applications in high-energy physics, we close
the presentation with a survey of some recent developments in analytic
solutions of hydrodynamics in Appendix~\ref{s:app:survey}.

\section{A rotating hydrodynamical solution}\label{s:solution}

Following Ref.~\cite{Csorgo:2013ksa}, we outline the hydrodynamical solution
valid for rotating expanding spheroids, which allows us to analytically
evaluate the observables.  The hydrodynamical problem is specified by the
continuity, Euler and energy equations:
\begin{eqnarray}
\partial_tn + \nabla\z{n\v{v}}    & = & 0 , \label{e:cont} \\
\partial_t\v{v} + \z{\v{v}\nabla}\v{v} & = &  -\nabla p / \z{m_0 n} , \label{e:Eu} \\
\partial_t\varepsilon + \nabla\z{\varepsilon\v{v}}  & = & - p\z{\nabla\v{v}} ,  \label{e:en}
\end{eqnarray}
where $n$ denotes the particle number density, $m_0$ is a the mass of an
individual particle (dominating the equation of state), 
$\v{v}$ stands for the non-relativistic (NR) flow velocity
field, $\varepsilon$ for the NR energy density and $p$ for the pressure. 
This set of equations
(\ref{e:cont}-\ref{e:en}) expresses five equations for six unknowns, and
are closed by some equation of state (EoS) that provides a relation
among pressure, energy and number density which relation can be
naturally expressed through their dependences on the temperature $T$. Just as in
Refs.~\cite{Csorgo:2001xm,Csorgo:2013ksa}, we choose a family of generalized
equations of state as:
\begin{eqnarray} 
p           = &  n T\,, \qquad
\varepsilon = &\kappa (T) n T .\label{e:eos} 
\end{eqnarray} 
This EoS allows us to study the solutions of NR hydrodynamical equations for any
temperature dependent ratio of pressure to energy density, $ p/\varepsilon =1/
\kappa (T) $, if the fireball evolution is also characterized by a conserved
particle number $n$. This approximation may be realistic, if in the final stages
of the hydrodynamical evolution, the hadrochemical reactions that may change
the particle types, are closed at temperatures above the kinetic freeze-out
temperature. Such an assumption is favoured by data~\cite{Adcox:2004mh,Adams:2005dq}
and is frequently used in the phenomenology of high energy heavy ion collisions.  For
a recent overview of the application of the concept of chemical freeze-out at or
near to the quark-hadron phase boundary, see e.g. ref.~\cite{Becattini:2012xb}.

The above EoS is thermodynamically consistent for any function $\kappa (T)$, as
was shown in Ref.~\cite{Csorgo:2001xm}.  The introduction of the function
$\kappa(T)$ as above contains several well-known special cases: a
non-relativistic ideal gas has $\kappa(T) = 3/2$, while for a gas of
relativistic massless particles one has $\kappa = 3$. Also, one can incorporate
a parametrization of the low temperature limit (after the hadrochemical
freeze-out) of lattice QCD equation of state into a suitable $\kappa(T)$
function. One also can model the change of the $p/\varepsilon$ ratio at a phase
transition from deconfined quark matter to hadronic matter, if one wants to
follow the time evolution from very high initial temperatures that corresponds
to deconfined quark matter, following the lines of ref.~\cite{Csorgo:2013ksa},
in which case the conserved charge density has to be replaced by the entropy
density $\sigma$ and the entalphy density 
$ \varepsilon  + p = \mu n + T \sigma$ 
will be dominated by the second term, $\varepsilon  + p \approx m_0 n$. 
 However, the hadronic observables evaluated in
the subsequent parts of this manuscript are formed in the final stages of the
hydrodynamical evolution, so we assume that the particle identity changing
hadrochemical reactions can be neglected close to the kinetic freeze-out
temperature and we proceed with the evaluation of the observables in this
approximation.  Note also, that one usually introduces the speed of sound as
$c_s^2 = dp/d\epsilon = 1/\kappa(T)$. Thus the above EoS allows for any
temperature dependent speed of sound, which can be taken either from
measurements or from fundamental calculations. Note also that the dynamics
of rotating and expanding fireballs was also considered for
even more general equations of state, where only the local conservation
of entropy density can be assumed but the chemical potential
is vanishing, so the entalphy density is dominated by the $ T
\sigma \gg \mu n$ term. This approximation, $\varepsilon + p \approx T\sigma$
is relevant   e.g. in case of  lattice QCD calculations for the equations of state
and the effect on the dynamics of fireball explosiveness has been evaluated and
discussed already in ref.  ~\cite{Csorgo:2013ksa}. Here we consider massive
particles driving the expansion as we focus on the dynamics around the
freeze-out time, that are relevant for the evaluation of the hadronic observables.

From now on, let us consider $n$, $\v{v}$ and $T$ as the independent unknown
functions, keeping in mind the caveats mentioned above and discussed also in
ref.~\cite{Csorgo:2013ksa}.  The hydrodynamical equations are solved, similarly
as it was done for the case of a non-rotating, NR ideal gas in
Ref.~\cite{Csorgo:2001xm}, by the following self-similar, ellipsoidally
symmetric density profile, and the corresponding velocity profile, that
describe the dynamics of a rotating and expanding
fireball~\cite{Csorgo:2013ksa}:
\begin{eqnarray}
T\z{\v{r},t}    &=& T(t), 	\label{e:Tansatz} \\
n\z{\v{r},t}    &=& n_0 \frac{V_0}{V} \exp\kz{-{\frac{r_x^2}{2X^2}} -{\frac{r_y^2}{2Y^2}} -{\frac{r_z^2}{2Z^2}}}, 
				\label{e:nansatz} \\
\v{v}\z{\v{r},t}&=& \z{\frac{\dot X}{X}r_x + \omega r_z\, ,\,\frac{\dot Y}{Y}r_y \,,\,\frac{\dot Z}{Z}r_z - \omega r_x}. 
				\label{e:vansatz}
\end{eqnarray}

Here $\omega = \omega(t)$ is the (time dependent) magnitude of the angular
velocity, and the scale parameters (the magnitudes of the ellipsoid axes) are
denoted by $(X,Y,Z)=(X(t),Y(t),Z(t))$, and the time dependence of the variables
is suppressed.  In addition, $\dot A$ stands for the time derivative of a time
dependent $A$ function: $\dot A = \frac{\mathrm{d}}{\mathrm{d t}}
A(t)$ and $\ddot A$ stands for the time derivative of $\dot A$.  The quantity
$V$ is a measure of the volume of the expanding system: $V = XYZ$. The initial
values of the temperature and this volume are then denoted by $T_0=T(t_0)$,
$V_0=V(t_0)$; while the $n_0$ quantity is a constant, corresponding to the
initial value of the density in the center.

Note that compared to Ref.~\cite{Csorgo:2013ksa}, we changed the labeling of
the axes: to correspond to the usual notation in heavy-ion physics, in this
paper we take the $r_z$ axis to be parallel with the beam axis, and the event
plane (the plane spanned by the impact parameter vector and the beam axis) is
assumed to be the $(r_x,r_z)$ plane. The rotation of the expanding fireball is
then assumed to be around the $r_y$ axis, because the initial angular momentum
points in this direction, while in ref.~\cite{Csorgo:2013ksa}, the  axis of
rotation was denoted by $r_z$.

In Ref.~\cite{Csorgo:2013ksa} it is shown that the above ansatz for the
temperature, velocity, and density fields indeed yields a solution to the
hydrodynamical equations if the following conditions are met. It turns out that
only ,,spheroidally symmetric'' solutions, ie. solutions with $X(t) = Z(t)
\equiv R(t)$ have been described so far in this way. The time evolution of the
radius parameters $X = Z = R$, the vertical scale parameter $Y$, the
temperature $T$ and the angular velocity $\omega$ are governed by the
following ordinary differential equations:
\bea
	X & = & Z \,  \equiv  \,  R , \\ \label{e:XZR}
	\dot X & = & \dot  Z \,  \equiv  \, \dot R , \\ \label{e:XZRdot}
	R\ddot R - R^2 \omega^2 &  = & \ddot{Y}Y \,  = \, \frac{T}{m_0} , \, 
		\\ \label{e:RYtime} 
	\dot T\frac{d}{dT}\z{T\kappa(T)} & = &  - T\frac{\dot V}{V} , \, 
		\\ \label{e:Ttime}
	\omega & = & \omega_0\frac{R_0^2}{R^2} . \label{e:omegatime}
\eea
The above formulae generalize the dynamical equations of
ref.~\cite{Csorgo:2001xm} for the case of non-vanishing initial angular
momentum, characterized here by the initial value of the angular velocity
$\omega_0$, and for spheroidal (X = Z = R) expansions.  For a vanishing value
of the initial angular momentum, $\omega_0 = 0$, the results of
ref.~\cite{Csorgo:2001xm} are reproduced on the level of the dynamical
equations, that characterize the expansion of the scale parameters $(X,Y,Z)$,
except the tilt angle, that was introduced in ref.~\cite{Csorgo:2001xm} to
characterize phenomenologically, with a time independent constant angle, the
effect of the rotation of the principal axis of the expanding ellipsoid in the
impact parameter plane with respect to the direction of the beam. In ref.~
\cite{Csorgo:2001xm} it was argued that due to the expanding nature of the
fireball, the moment of inertia increases for late stages of the expansion, so
the angular velocity slows down due to the conservation of angular momentum,
hence in the final stages of the expansion approximately a constant value of
the tilt angle can be used to evaluate the observables. 

By now, we are able to solve the dynamical equations that describe the rotation
of the fireball in the impact parameter plane, and so the tilt angle $\vartheta
= \vartheta(t) $ and the angular velocity $\omega = \dot\theta$ 
in this manuscript are both time dependent functions, that can be
evaluated for the expanding and rotating ellipsoid by integrating the time
dependent angular velocity as follows:
\begin{equation}  
\vartheta = \theta_0 + \int_{t_0}^t \m{d}t^\prime \, \omega(t^\prime) . \label{e:omegatheta} \\
\end{equation}

Following Ref.~\cite{Csorgo:2001xm}, the temperature equation \Eq{e:Ttime} can
also be integrated as 

\begin{equation} 
\frac{V_0}{V} = \exp\left[\kappa(T) -
\kappa(T_0)\right] \exp\int_{T_0}^T \frac{dT^\prime}{T^\prime}
\kappa(T^\prime)\,, 
\end{equation} 

and in the case of temperature independent
$\kappa$, ie. $\kappa(T) = \kappa = const$, this relation can be expressed in
the simple and instructive way as 

\begin{equation}
 T = T_0 \left(\frac{V_0}{V}\right)^{1/\kappa}.  
\label{e:T32}
\end{equation} 

This equation is formally the same as the temperature law for adiabatic expansions of
homogeneous gases of volume V, where $T V^{\gamma-1}= const$, and 
the adiabatic index is introduced as $\gamma = (f +2) / f$, 
where $f$ stands for the elementary degrees of
freedom. Identifying the adiabatic index $\gamma$ with $ 1 + 1/\kappa $ one
obtains the relation $f = 2 \kappa$, so one can relate the coefficent between
the energy density and the pressure to the number of degrees of freedom in the
manner usual in thermodynamics of ideal gases. 
In our particular case, $\kappa \equiv \kappa(T) $ is 
a temperature dependent function, and so are the effective number 
of degrees of freedom, $f \equiv 2 \kappa(T)$.  
Let us also emphasize, that  we discuss
an exact solution of a rotating and expanding hydrodynamical system, where the
density, the velocity and the pressure fields are all functions of space and
time, so the emergence of the law of adiabatic expansions of thermodynamics is
a beautiful by-product of the fact that we were able to reduce 
the complicated system of partial differential equations of hydrodynamics
to a system of coupled, nonlinear but ordinary differential equations.
In this particular case, the occurrence of the law of adiabatic expansions
of thermodynamics corresponds precisely to the adiabatic or perfect fluid 
nature of the exploding and rotating Gaussian fireball under investigation.

The time evolution of the principal axes $R$ and $Y$ are given by
\Eq{e:RYtime}, which can be understood as a classical motion of a point
particle with mass $m$ in a non-central potential. The corresponding
Hamiltonian $H\z{P_R,P_Y,R,Y}$ takes a simple form in the $\kappa = const$
case\footnote{Ref.~\cite{Csorgo:2013ksa} details the Hamiltonians for other
choices of the $\kappa(T)$ function, or for the case when there is no conserved
charge $n$.}:
\begin{equation}
H = \frac{P_R^2+2P_Y^2}{4m_0} + \frac{m_0\omega_0^2R_0^4}{R^2} + T_0\kappa \z{\frac{R_0^2Y_0}{R^2Y}}^{1/\kappa} ,
\label{e:Hamilton}
\end{equation}
from which $P_R = 2m_0\dot R$, $P_Y = m_0\dot Y$ follows. One can verify that the
Hamiltonian equations of motion are indeed the ones in \Eq{e:RYtime}. A first
integral of the motion is simply the conservation of the energy $E$:
\[
\frac{E}{m_0} = 
\dot R^2+\frac{\dot Y^2}{2} + 
\frac{\omega_0^2R_0^4}{R^2} + 
\frac{T_0\kappa}{m} \z{\frac{R_0^2Y_0}{R^2Y}}^{1/\kappa} = \m{const} .
\]
This equation, with the help of the form of the solutions for the temperature $T$
and the angular momentum $\omega$ can also be rewritten to the following, 
rather intuitive form:
\be
E = m_0 \left( \dot R^2+\frac{\dot Y^2}{2} + \omega^2 R^2 \right) + \kappa T ,
\ee 
which indicates that the dependence of the temperature on the volume $V = R^2 Z$ plays a role of an effective, non-central potential in the corresponding Hamiltonian problem.  If the special case $\kappa = 3/2 $ is considered (this
is the case of a classical ideal gas), the Hamiltonian problem can be solved
using quadratures, see Appendix~\ref{s:app:H} for the details. An interesting
result is that in this case the time dependence of the $2R^2+Y^2$ quantity can
be expressed explicitely for arbitrary initial conditions:
\be
2R^2+Y^2  =  2\z{\dot R_0 t +R_0}^2 + \z{\dot Y_0 t + Y_0}^2 + 
	 \z{\frac{3T_0}{m_0}+2\omega_0^2R_0^2}t^2 .  \label{e:R2Y2}
\ee
This formula generalizes the earlier result of ref.~\cite{Csorgo:2001xm}
for the case of non-vanishing initial angular momentum, characterized here
by the initial value of the angular velocity $\omega_0$. For a vanishing
value of $\omega_0$, and for the spheroidal expansions of
$X = Z =R$, the earlier results of ref.~\cite{Csorgo:2001xm} are reproduced. 

For other choices of $\kappa$, one has to resort to numerical solutions of
Eq.~\ref{e:RYtime}, the equation of motion. Such solution can be easily found 
numerically for any initial conditions $R_0$, $Y_0$, $\dot R_0$,
$\dot Y_0$, $\omega_0$ with the presently available tools of desktop mathematics 
(e.g. Mapple, MathLab or Mathematica).

\section{Observables from the new solution}\label{s:obs}

In the previous section we have seen how the hydrodynamical equations of a
rotating and spheroidally symmetric, expanding fireball can be reduced to an
easily solvable system of ordinary differential equations and we have also
derived some new first integrals of the motion. To obtain some physically
interesting particular solutions, one needs to specify the equations of state,
the initial conditions and the freeze-out conditions.
In this manner, one can can easily investigate the
effect of rotation on the time evolution of the system.  
In this section we illustrate this dynamics with some examples and proceed
also with the analytic evaluation of the observables from the dynamical solutions,
then we also illustrate these analytic results with some numerical examples.

For the sake of illustration, we take the initial conditions to be that of a
sphere, with $X_0 = Z_0 \equiv R_0 = 5$ fm, $Y_0 = 5$ fm, and $T_0 = 350$ MeV.
For the sake of simplicity, we switch off the effects of initial radial flows,
$\dot X_0 = \dot Z_0 = \dot R_0 = 0$,  as in these numerical examples we focus
on the effects of initial angular momentum that can be demonstrated more
clearly if the interference with radial flow effects (inherent in the formulas)
are switched off for the purpose of these numerical examples.  We compare three
cases, each of different initial angular momentum, corresponding to  $\omega_0$
values increasing gradually from nearly zero to a more realistic value of 0.1 fm/c. 
For the sake of
these illustrations, we take the simple $\kappa = 3/2$ value for the EoS and
expect qualitatively similar results for other choices of the $\kappa(T) $
function and we set the $m_0$ parameter of the dynamical equations 
of motion to be the proton mass, $m_0 = 938$ MeV, as in Ref~\cite{Csernai:2014hva}
the dynamics of the numerical solution of a 1+3d fireball hydrodynamical problem
with lattice QCD equation of state was well approximated with our simple
analytic fireball hydrodynamic solution of ~\cite{Csorgo:2013ksa},
used to model the dynamics also in our present manuscript.

\begin{figure}%
\begin{center}
\includegraphics[width=0.75\columnwidth]{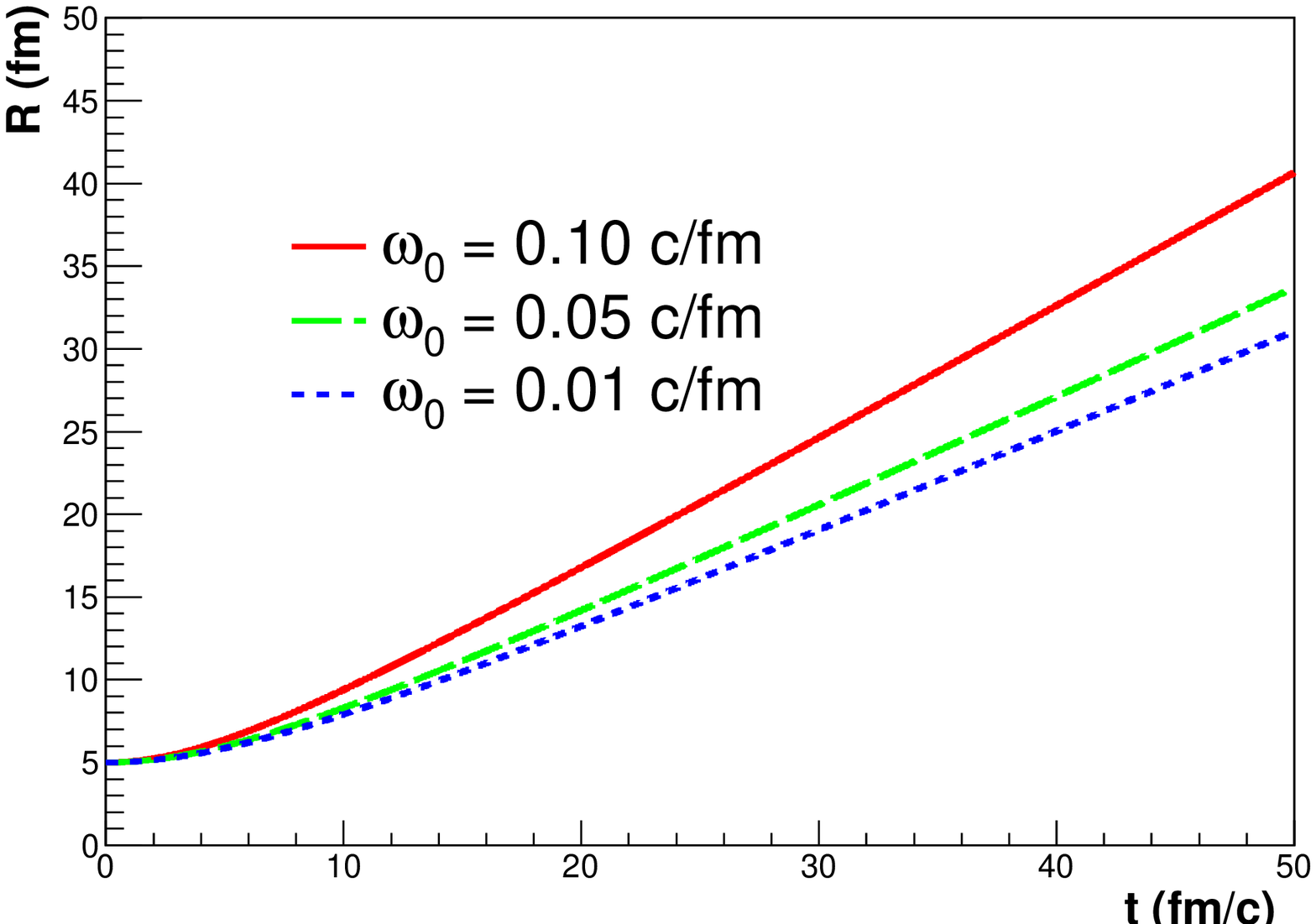}\\
\includegraphics[width=0.75\columnwidth]{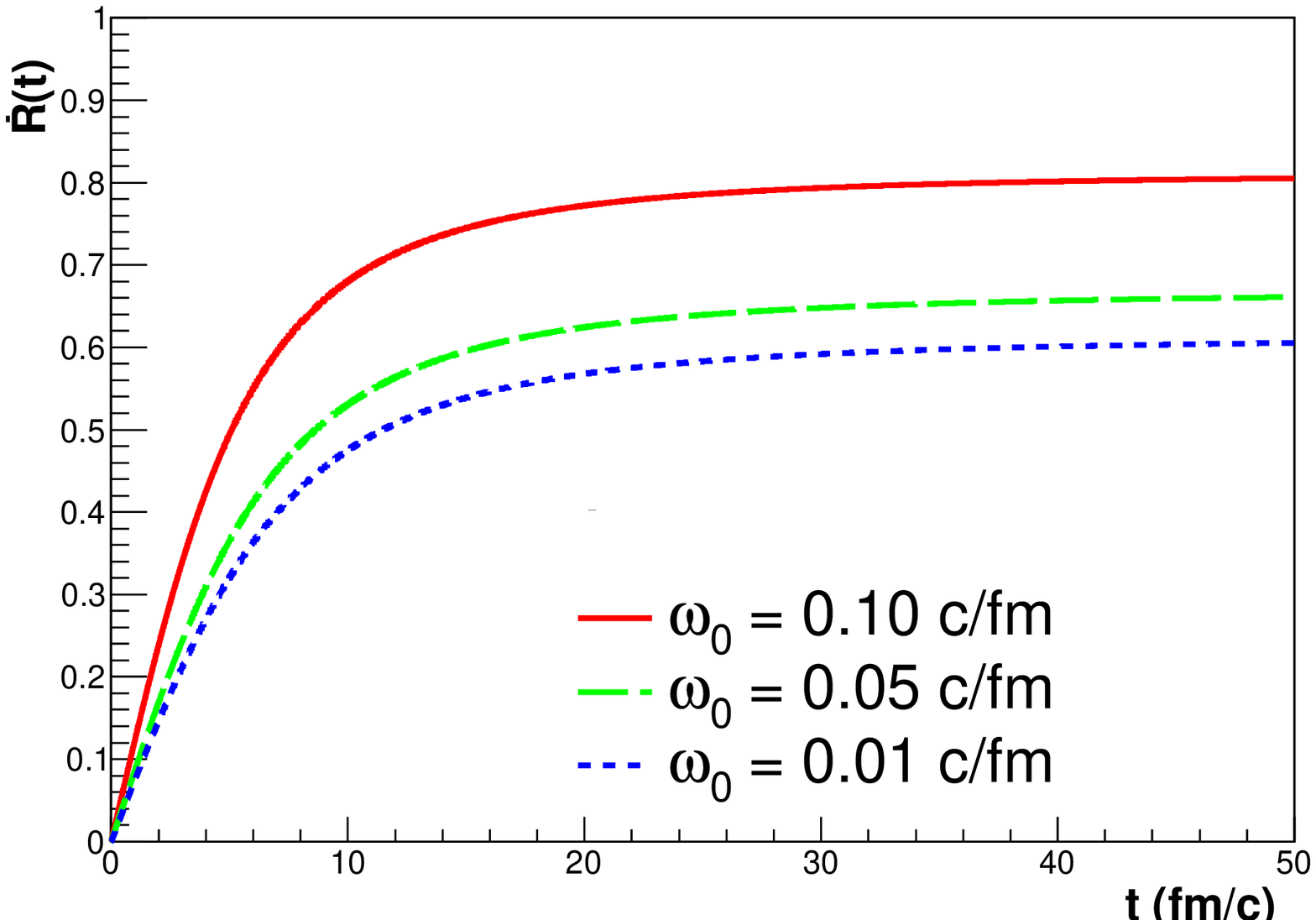}\\
\caption{Time evolution of the axes $X(t)=Z(t)\equiv R(t)$ (upper panel) and
the corresponding velocity, $\dot R(t)$ (lower panel) for our hydrodynamical
model in three cases of $\omega_0$. 
Initial conditions are: $R_0 = 5$ fm, $Y_0
= 5$ fm, $\dot R = \dot Y = 0$, $T_0 = 350 $ MeV. One sees that increased initial
angular momentum leads to increased, more violent explosion in the radial direction.}%
\label{f:R}%
\end{center}
\end{figure}

\begin{figure}%
\begin{center}
\includegraphics[width=0.75\columnwidth]{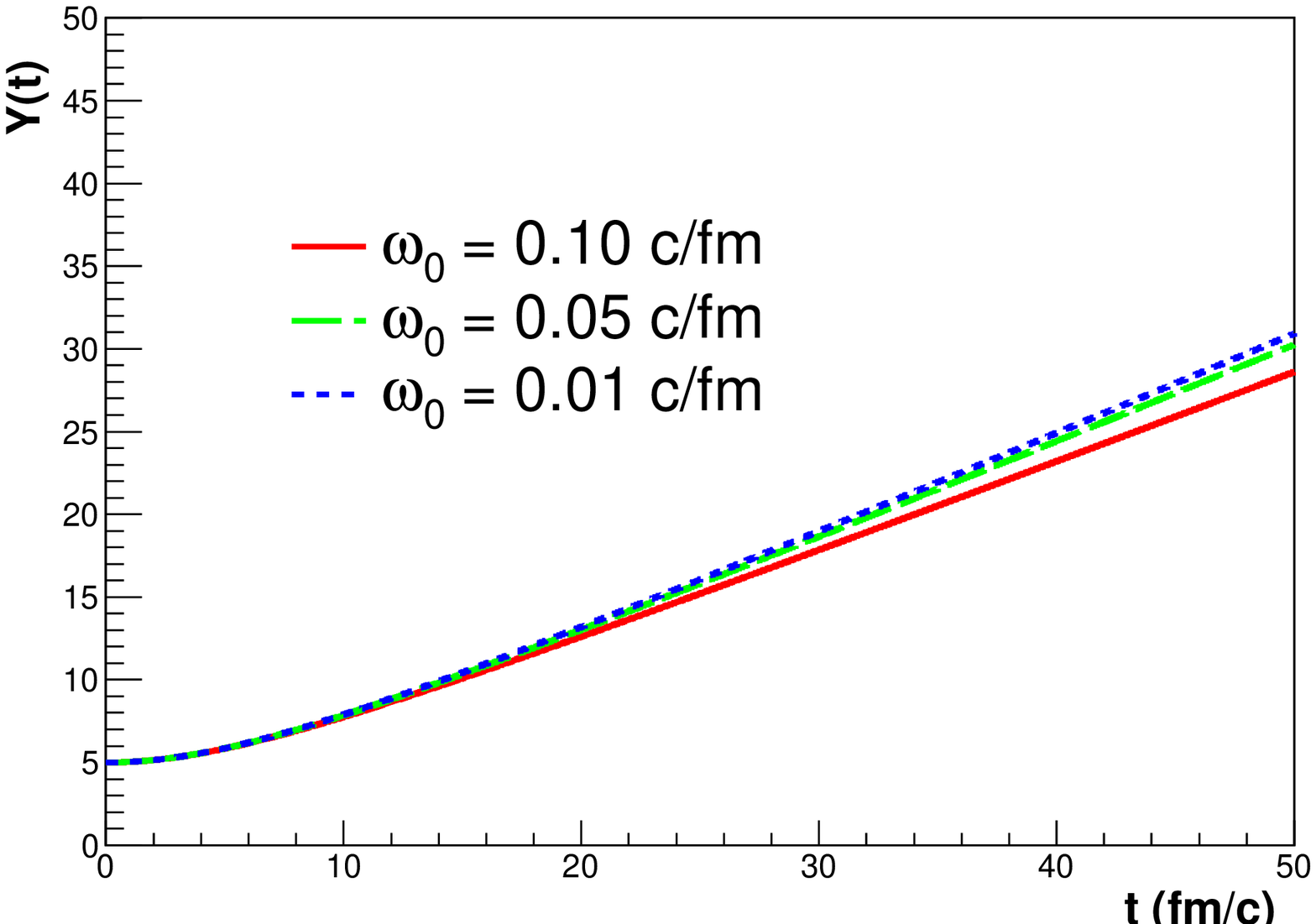}\\
\includegraphics[width=0.75\columnwidth]{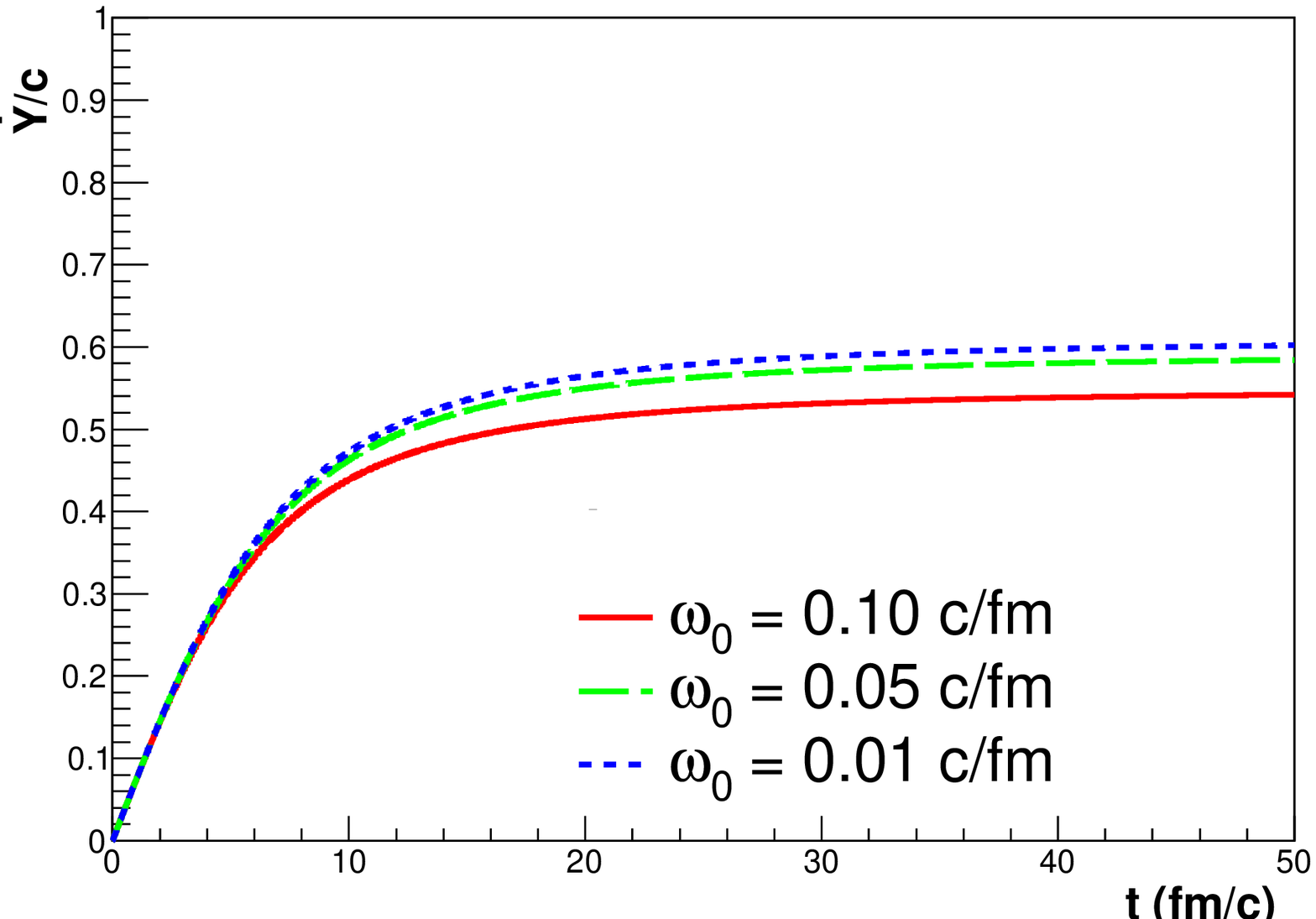}\\
\caption{Time evolution of the $Y(t)$ axis (upper panel) and $\dot Y(t)$ (lower
panel) for our hydrodynamical model in three cases of $\omega_0$.  
Initial conditions are the same as in Fig.~\ref{f:R}. 
The more violent radial expansion related to increased values of initial
angular momentum results in decreased expansion in the $Y$ direction, due to the
faster cooling and the correspondingly faster decrease of the pressure of the fireball.
}%
\label{f:Y}%
\end{center}
\end{figure}

\begin{figure}%
\begin{center}
\includegraphics[width=0.75\columnwidth]{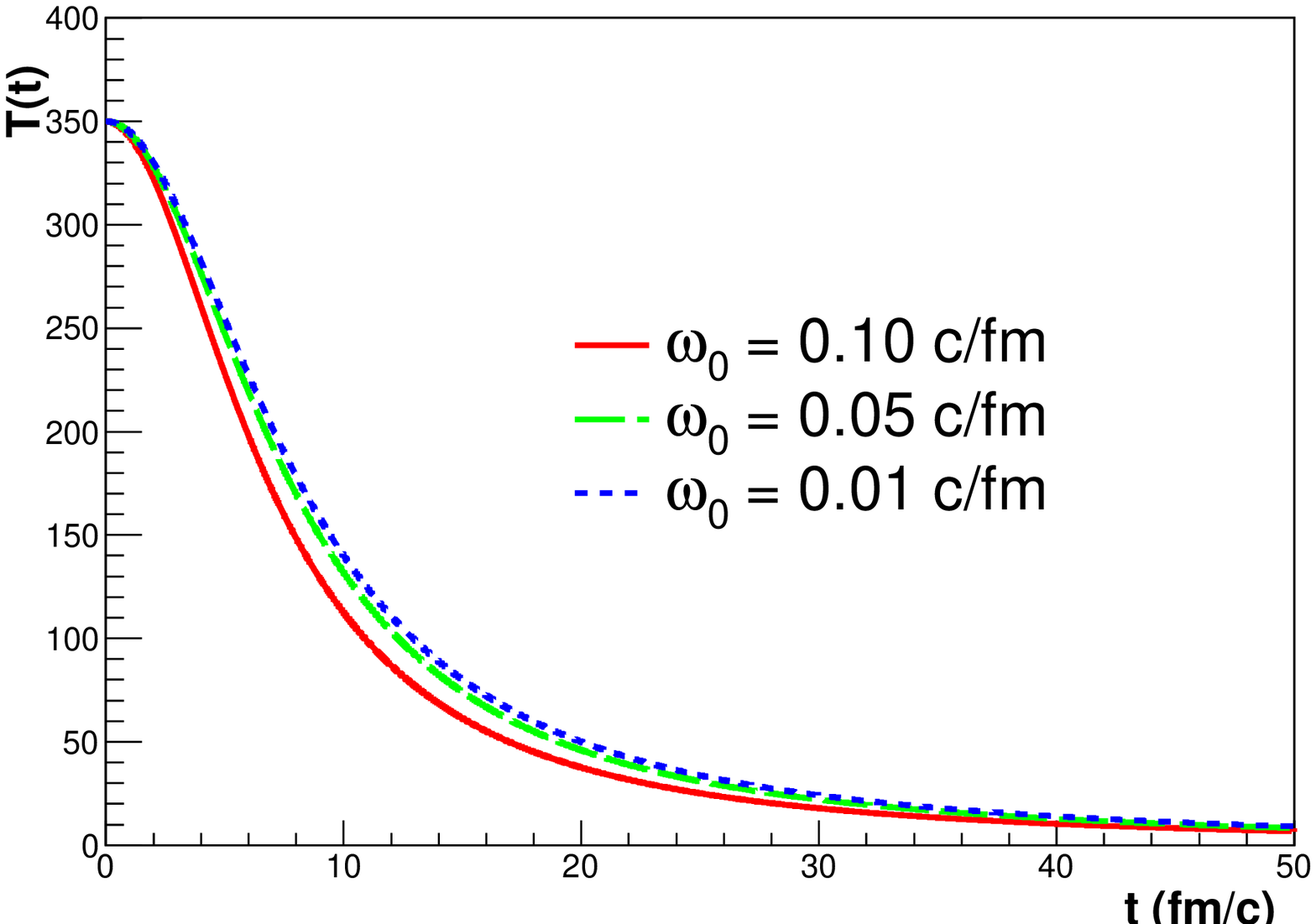}\\
\includegraphics[width=0.75\columnwidth]{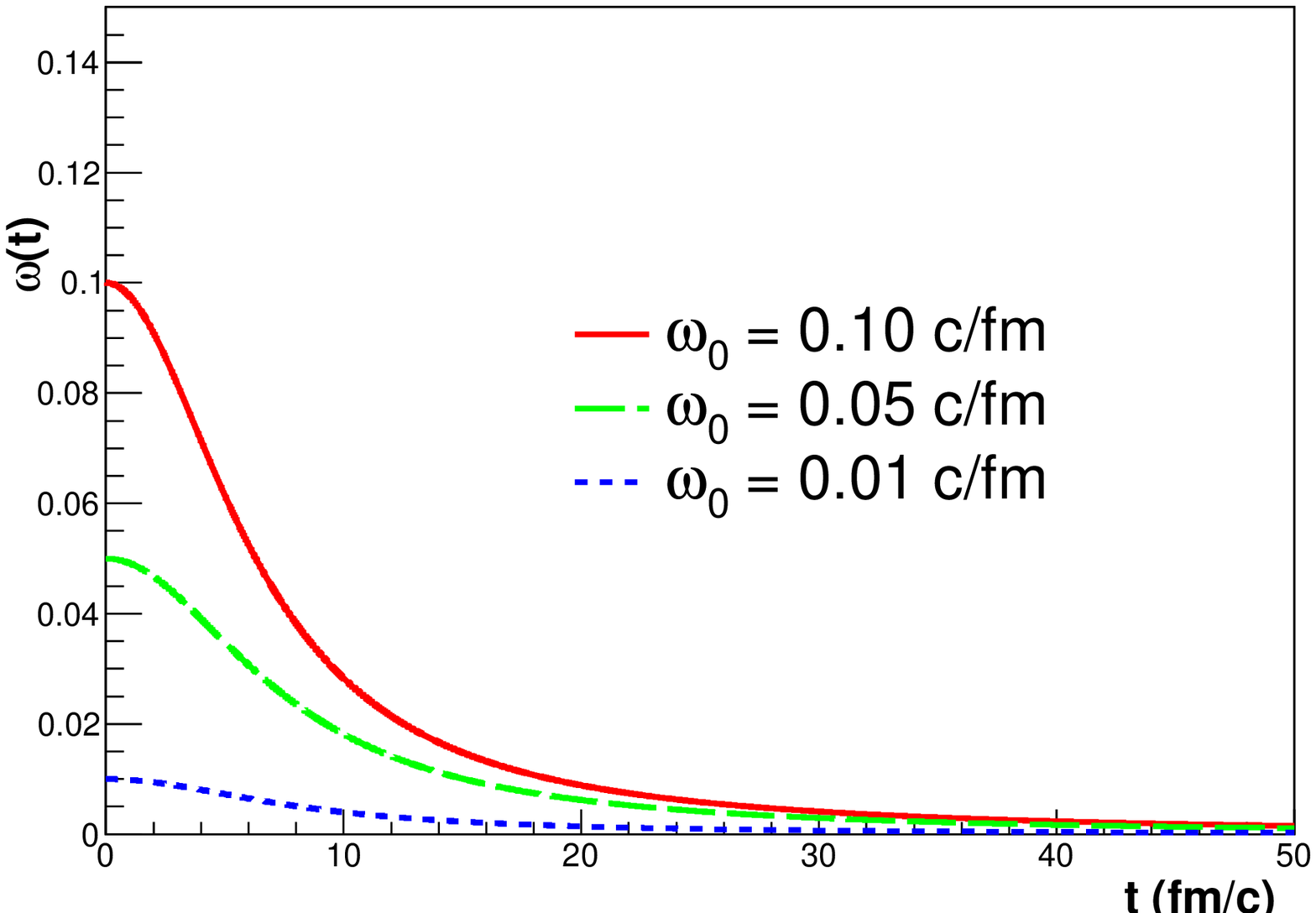}\\ 
\caption{ 
Time evolution of the temperature (upper panel) and the angular velocity (lower
panel) for our hydrodynamical model in three cases of $\omega_0$. In this
demonstrational case, $\kappa = 3/2$, so the temperature is directly related to
the geometrical size of the fireball through \Eq{e:T32}.  Initial conditions
are the same as in Fig.~\ref{f:R}.  Larger initial angular momentum leads to
faster radial expansion, that causes faster cooling.  In this class of rotating
and expanding, spheroidal fireball solutons, the time evolution of the $r_y$
component of the vorticity vector, $\mathbf{\omega}_y(t)$ equals with twice the
angular velocity $\omega(t)$ indicated on the lower panel, due to
eq.~(\ref{e:vorticity}).  
}  
\label{f:TOW}  
\end{center} 
\end{figure}

Figs.~\ref{f:R}, \ref{f:Y}, and \ref{f:TOW} show the time evolution of the
$X(t)=Z(t)\equiv R(t)$ and the $Y(t)$ axes, the temperature and the angular
velocity, respectively. One indeed sees, what one expects based on the analytic
structure of the dynamical equations of eqs.~(\ref{e:XZR}-\ref{e:omegatime}),
namely  that increased initial angular
momentum leads to an accelerated expansion in the $R$ direction, and faster
cooling, and this leads to a slightly decreased expansion in the $Y$ direction.

In order to evaluate the measurable quantities from the hydrodynamical solution,
a further specification needs to be made, namely, the condition of the
freeze-out (that sets the end of the hydrodynamical evolution) has to be
stated.  Let us introduce the $f$ subscript to indicate that the quantity is to be 
taken at the freeze-out time, $t_f$. 
Here we assume that when the temperature reaches a given $T_f $  value, a
sudden freeze-out happens, and identical particles with mass $m$ are produced.
We evaluate some of the observables for pions, kaons and protons or anti-protons,
where for the observed particles we use  $m = 140, 494$ and $938$ MeV, respectively.

The $T_f \equiv T(t_f)$ 
temperature is reached everywhere at the same time $t_f$ in the considered
class of solutions, as seen from \Eq{e:Tansatz}. The emission function can be
written as
\begin{equation}
S\z{t,\v{r},\v{p}} = \frac{n}{\z{2\pi mT}^{3/2}}\exp\left\{- \frac{\z{\v{p}-m\v{v}}^2}{2mT}\right\}\delta\z{t-t_f} ,
\label{e:Sdef}
\end{equation}
where the space and time dependence of the $\v{v}$, $T$, $n$ fields was
suppressed in the notation. This emission function is normalized so that its
integral over $\v{p}$ at a given point $\v{r}$ indeed yields the number
density, $n$ at that point.  We also assume that at freeze-out, the equation of
state tends to that of an ideal gas: $\lim_{T\rightarrow T_f} \kappa(T) 
= 3/2$.  Using our forms for the $\v{v}$, $T$, $n$ fields specified above in
Eqs.~(\ref{e:vansatz}), (\ref{e:nansatz}), and (\ref{e:Tansatz}), one finds
that the emission function has a multi-variate Gaussian form, and the integrals
can performed analytically. 


The effect of the rotation on the expansion dynamics is illustrated on Figures
~\ref{f:R},~\ref{f:Y},~\ref{f:TOW}, that indicate the time evolution of an initially spherical, rotating volume with the same initial geometry $X_0 = Y_0 = Z_0$, without initial 
radial flows, $\dot X_0 = \dot Y_0 = \dot Z_0 = 0$ and the
same initial temperature $T_0$ but with various initial angular momentum 
$\omega_0$ increasing from 0 to 0.1 . 
On Figure ~\ref{f:R} one sees that the expansion becomes yields more radial flow 
with increasing initial angular momentum.
Figure~\ref{f:Y} indicates that the more violent radial expansion in the cases with bigger angular momentum causes slower expansion in the $Y$ direction.
Figure~\ref{f:TOW} indicates, that the temperature of the fireball cools faster with
more rotation, which is a direct consequence of the faster increase of the volume
of the fireball with increasing initial angular momentum.
In our calculations of the observables, we took $T_f =
140$ MeV freeze-out temperature (see Section~\ref{s:obs}), which is reached
after approx.\ 8-10 fm$/c$ time, depending weakly on the initial angular momentum,
as illustrated on Fig.~\ref{f:TOW}.

Let us introduce the following notation:
\begin{eqnarray}
T_x & = & T_f + m \z{\dot X_f^2+{ \omega_f^2 Z_f^2}} \, 
 	= \, T_f + m \z{\dot R_f^2+{ \omega_f^2 R_f^2}} 
	, \, \label{e:txp}\\
T_y & = & T_f + m \, \dot Y_f^2 , \, \label{e:typ}\\
T_z & = & T_f + m \z{\dot Z_f^2+{ \omega_f^2 X_f^2}} 
 	= \, T_f + m \z{\dot R_f^2+{ \omega_f^2 R_f^2}} 
	.  \label{e:tzp}
\end{eqnarray}
We chose the $x$, $y$, $z$ subscripts to denote the principal directions and
also the $X$, $Y$, $Z$ notation to denote the size of the axis of the expanding
spheroid in the corresponding principal directions for the sake of clarity and
reminiscence to earlier results, but remember that in the present solution $X =
Z$, so $\dot X = \dot Z$, and $T_x = T_z$.  
Based on the effect of the rotation on the expansion dynamics
as illustrated on Figures~\ref{f:R},\ref{f:Y},\ref{f:TOW} it 
is plausible to assume that the
initial conditions imply $T_x = T_z > T_y$ in practical cases, when the initial
radial flows are expected to be small, $\dot X \approx \dot Z \approx 0$.

Rotational fluids are frequently characterized in terms of their vorticity vector, 
defined as 
$\mathbf{\omega}(\mathbf{r},t) = \nabla \times \mathbf{v}(\mathbf{r},t)$.
For this class of rotating and expanding, spheroidal fireball solutions,
the vorticity vector has been found~\cite{Csorgo:2013ksa}
to be spatially homogeneous, pointing to the axis of rotation, and directly proportional to the value of the angular velocity of the fluid,
the scalar function $\omega(t)$, that we recapitulate here for the sake of completeness. In the present manuscript the principal axis of the rotation corresponds to the 
$r_y$ direction, while in ref.~\cite{Csorgo:2013ksa}
the axis of rotation was pointing to the $r_z$ direction, hence in the notation 
of the current manuscript the vorticity for this family of rotating and expanding
exact solutions of fireball hydrodynamics reads as
\begin{equation}
\mathbf{\omega}(\mathbf{r},t) = \nabla \times \mathbf{v}(\mathbf{r},t)
		= (0, 2 \omega(t), 0).
		\label{e:vorticity}
\end{equation}
Due to this reason, the time evolution of the angular velocity 
$\omega \equiv \omega(t)$,
indicated on Fig.~\ref{f:TOW} equals to the time evolution of 
half of the $r_y$ component of the spatially homogeneous vorticity vector,
$\mathbf{\omega}_y \equiv \mathbf{\omega}_y (t)$:
\begin{equation}
		 \omega = \frac{1}{2} \mathbf{\omega}_y .
		\label{e:vorticity-fun}
\end{equation}

In what follows, we demonstrate that the parameters $T_x$ and $T_y$
introduced  in eqs.  (\ref{e:txp}-\ref{e:tzp}) in this manuscript
play a very similar role to that of the analogous variables
introduced in Ref.~\cite{Csorgo:2001xm}. 
There, the dynamics of the rotation was not taken into account,
consequently the expressions of $T_x$ and $T_y$ lacked the term corresponding to the
rotational energy, $R^2\omega^2$ (with $X=Z=R$), but these 
terms are clearly identified 
in the present manuscript.
This way we prove, that in this class of rotating solutions of fireball
hydrodynamics the effect of rotation 
enters the final state in a  rather straightforward way: 
the effects that were usually associated with that of radial flows 
need to be re-considered to take the rotation effects also into account.

The presence of rotation changes the time evolution of the
system and thus for a given initial condition, it influences the final state in
an involved way. It is thus rather reassuring to find the simple forms of
the rotational terms in the expression of the effective
temperatures $T_x=T_z$ and $T_y$, as a manifestation of rotation.

\subsection{Single particle spectrum}

The single-particle spectrum is basically the integral of $S(\v{r},\v{p})$ over
the spatial coordinates. The result is
\begin{eqnarray}
E\frac{dn}{d^3\v{p}} & \propto & 
E\exp\kz{-\frac{p_x^2}{2 m T_x} -\frac{p_y^2}{2 m T_y} -\frac{p_z^2}{2 m T_z}} . 
	\label{e:ellsp}
\end{eqnarray}
Here $E$ is the energy of the particle. In our case, $T_x = T_z$, since $X_f =
Z_f$. In this case the spectrum is rotationally invariant in the $x$--$z$
plane, so the observables calculated in this way are automatically valid in the
laboratory frame as well as the rotated frame of the fluid.  However, in the
more general, yet to be explored case of a rotating and expanding ellipsoid
with three different axes, this statement is generally not true: in that case,
the observables calculated in the rest frame of the collision will differ from
those measured in the proper frame of the ellipsoid at freeze-out.

The spectrum seen in Eq.~(\ref{e:ellsp}) generates the following azimuthally
averaged single-particle spectrum ($p_t$ stands for transverse momentum):

\begin{eqnarray}
\frac{dn}{2\pi p_t dp_t dp_z} 
	& \propto &
	\exp\left(-\frac{p_t^2}{ 2 m T_{\rm eff}} -\frac{p_z^2}{ 2 m T_z}\right)
	I_0(w) , \label{e:ellsptr} \\
 \frac{1}{T_{\rm eff}} 
	& = &
	 \frac{1}{2}\z{\frac{1}{T_x} + \frac{1}{T_y} }, \label{e:Teffdef} \\
       w  
	& = &
	 \frac{p_t^2}{4 m}\z{ \frac{1}{T_y} - \frac{1}{T_x} } . \label{e:www}
\end{eqnarray}

Here and in what follows, 
the modified Bessel function of order $n$ is denoted by $I_n\z{w}$, which  
is defined as $I_n(w) = \rec\pi\int_0^\pi \m dz \cos(n z) \exp\sz{w\cos(z)}$.

Just as we did for the time evolution of the system itself (see
Figs.~\ref{f:R}--\ref{f:TOW}), we demonstrate the observables in the
illustrative $X_0 = Y_0 = Z_0 = 5$ fm, $\dot X_0 = \dot Y_0 = \dot Z_0 = 0$,
and $T_0 = 350$ MeV case, taken with three different values of initial angular
velocity $\omega_0$, to display the effect of rotation on them. 


In a relativistic generalization or extension, 
the mass dependence of the
slope parameters transforms into the $m_T$ (transverse mass) dependence, 
see e.g. Refs.~\cite{Csanad:2003qa},\cite{Csorgo:1995bi},
so our
result on the $m$ dependence can be taken as a preliminary suggestion on the
$m_T$ dependence what one might get in a more realistic relativistic approach.
In our calculations, from now on we take the freeze-out temperature $T_f$ to be the
pion mass, 140 MeV.


\begin{figure}%
\begin{center}
\includegraphics[width=0.5\columnwidth]{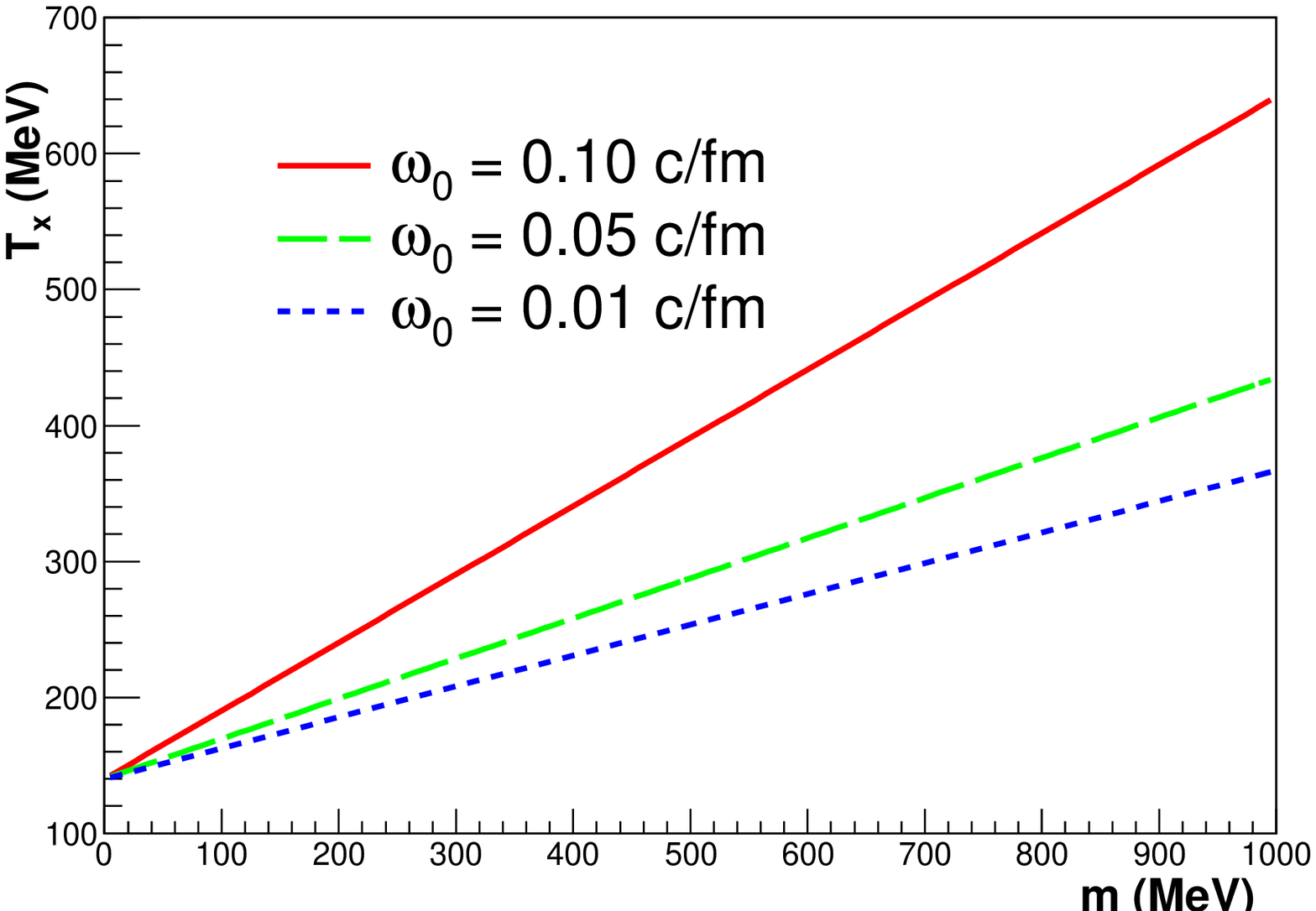}\\
\includegraphics[width=0.5\columnwidth]{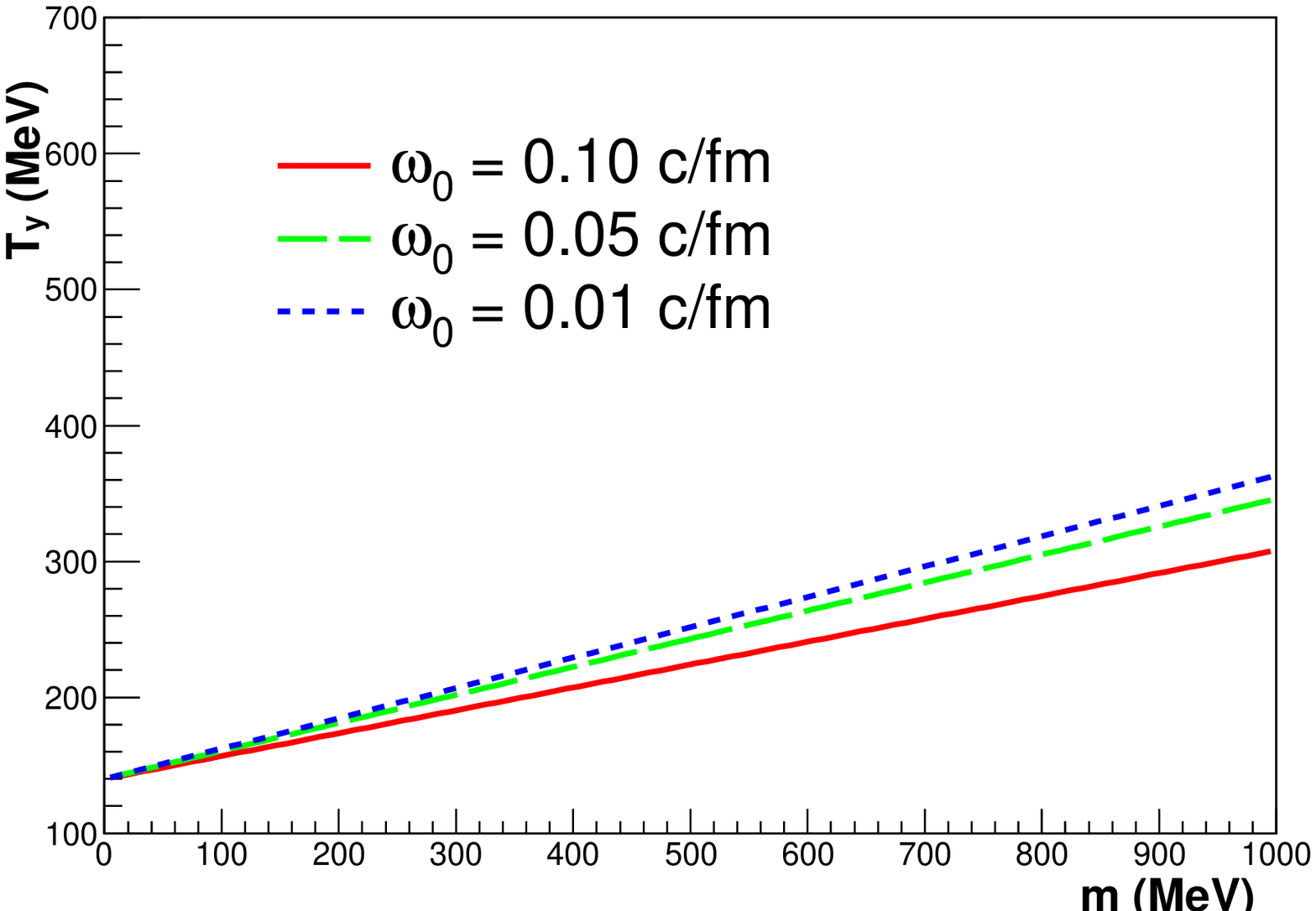}\\
\includegraphics[width=0.5\columnwidth]{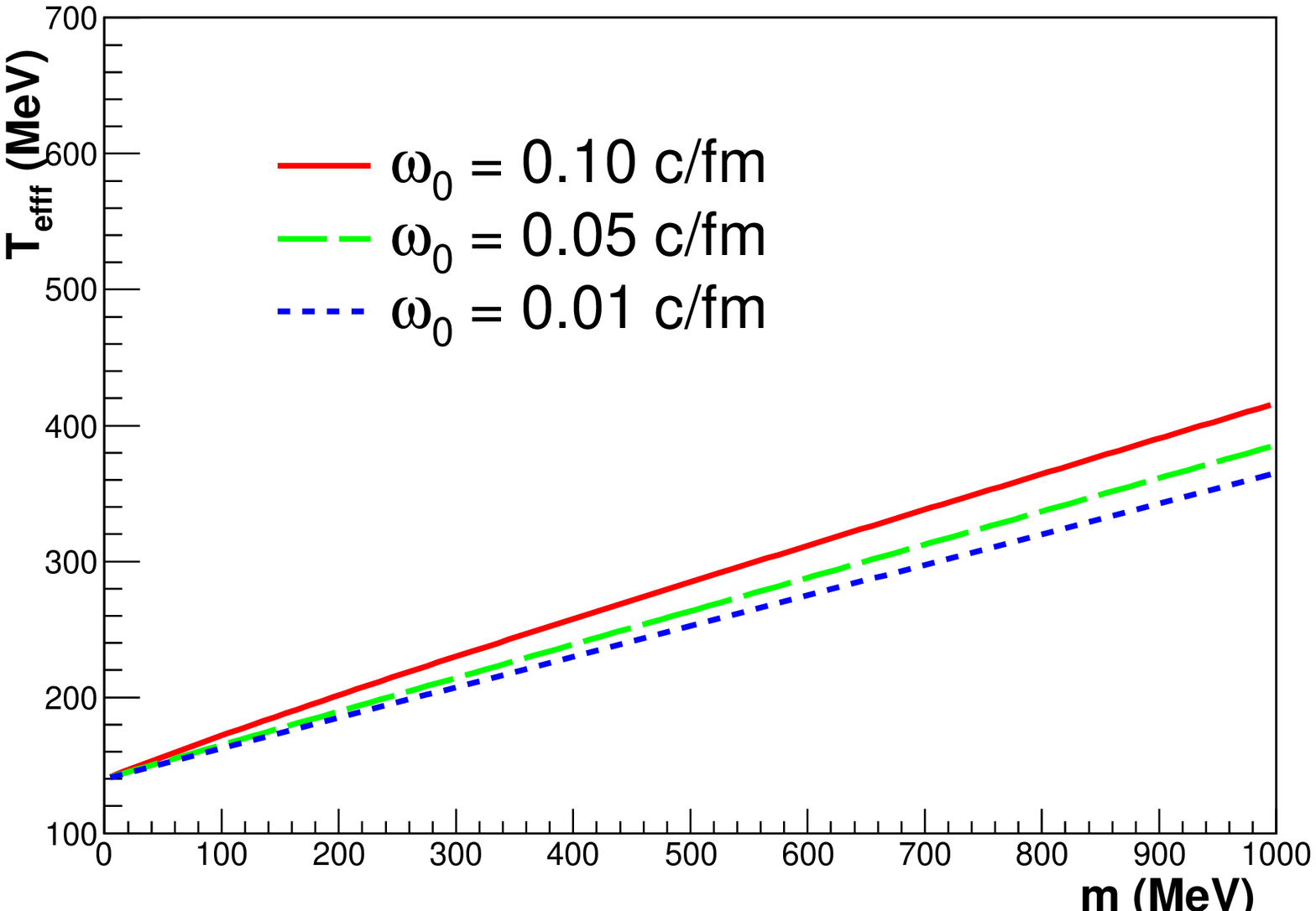}\\
\caption{
The slope parameters $T_x = T_z$ (upper panel), $T_y$ (middle panel) and the
effective temperature introduced in \Eq{e:Teffdef}, $T_{\rm eff}$ (lower
panel), taken at freeze-out (when $T$ reaches $T_f = 140$ MeV), as a function
of particle mass $m$. As before, three cases of $\omega_0$ are selected for
the purpose of demonstration of the analytically obtained formulas.
  Initial conditions are the same as in Fig.~\ref{f:R}.  
Larger initial angular momentum creates additional terms that add to the
radial flow effects, thus larger initial angular momentum leads to 
higher slope parameters or effective temperatures in the single particle spectrum..
}%
\label{f:fTxye}%
\end{center}
\end{figure}

The azimuthal dependence of the single-particle spectrum can be encoded in the
Fourier components of it, these are called the flow coefficients, and denoted
by $v_n$. They are defined as
\begin{equation}
\frac{dn}{dp_z p_t dp_t d\phi} 
	= 
	\frac{dn}{2 \pi dp_z p_t dp_t} 
		\left[1 + 2 \sum_{n=1}^{\infty} v_n \cos(n\phi)\right]. 
\label{e:harm}
\end{equation}
Here $v_1$ is called the directed flow, $v_2$ the elliptic flow and $v_3$ the
third flow. It is important to note that in our simple model all the event
planes coincide and that is where we set the zero of the azimuthal angle,
that's why the sinusoidal terms do not appear in \Eq{e:harm}.

It turns out that in our model the transverse- and longitudinal momentum
dependence of the $v_n$ flow components can be written in terms of the $w$
scaling variable introduced earlier in \Eq{e:Teffdef}. Simple calculation leads
now to
\begin{equation}
v_{2n+1} = 0, \quad v_{2n} = \frac{I_n(w)}{I_0(w)} . \label{e:vn} 
\end{equation}
The odd components of the Fourier decomposition vanish. This is an artefact due
to the spheroidal symmetry of the flow profile.  In a more general case of
non-spheroidal rotating expansion of an ellipsoid mentioned earlier, after
\Eq{e:ellsp}, but not detailed in this manuscript, the observable angular tilt
of the fireball would imply non-vanishing odd azimuthal moments. However, the
even components mirror the effect of rotation as well: the $w$ variable, which
governs the dependence of the flow components on the kinematic
variables\footnote{One should clearly distinquish the angular velocity of the
rotiating fluid $\omega= \omega(t)$ not only from the vorticity vector,
$\mathbf{\omega}(\mathbf{r},t) = \nabla \times \mathbf{v}(\mathbf{r},t)$,
but also from from the scaling variable of the elliptic flow, 
denoted here by $w$ following the notation introduced already in 
ref.~\cite{Csorgo:2001xm}.  
The latter scaling $w$
variable enters in this manuscript
as the argument of $I_n\z{w}$.}, and it increases compared to the
non-rotating case of Ref.~\cite{Csorgo:2001xm}, corresponding to the increase
of the elliptic flow due to the effect of rotation.

\begin{figure}%
\begin{center}
\includegraphics[width=0.75\columnwidth]{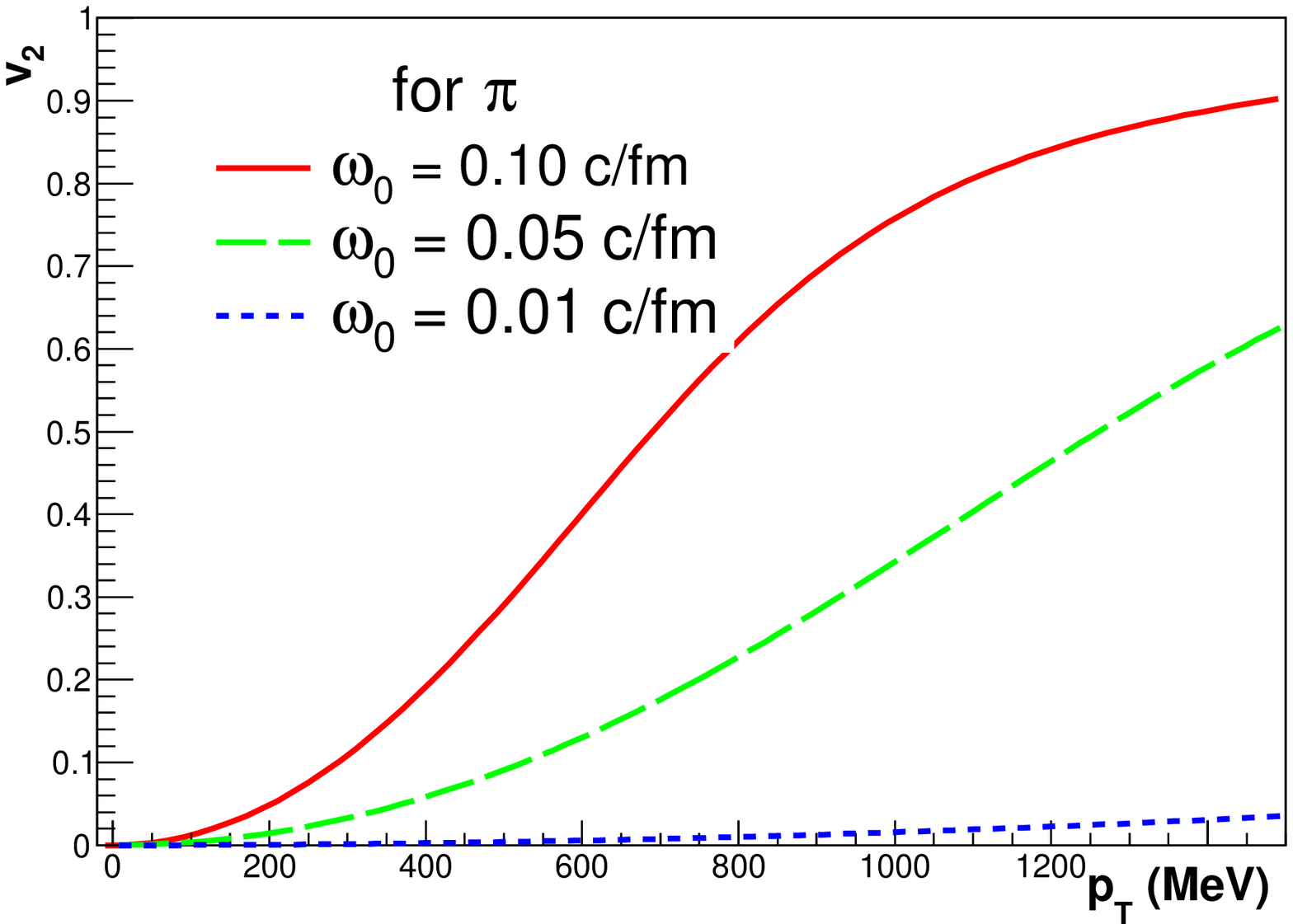}\\
\includegraphics[width=0.75\columnwidth]{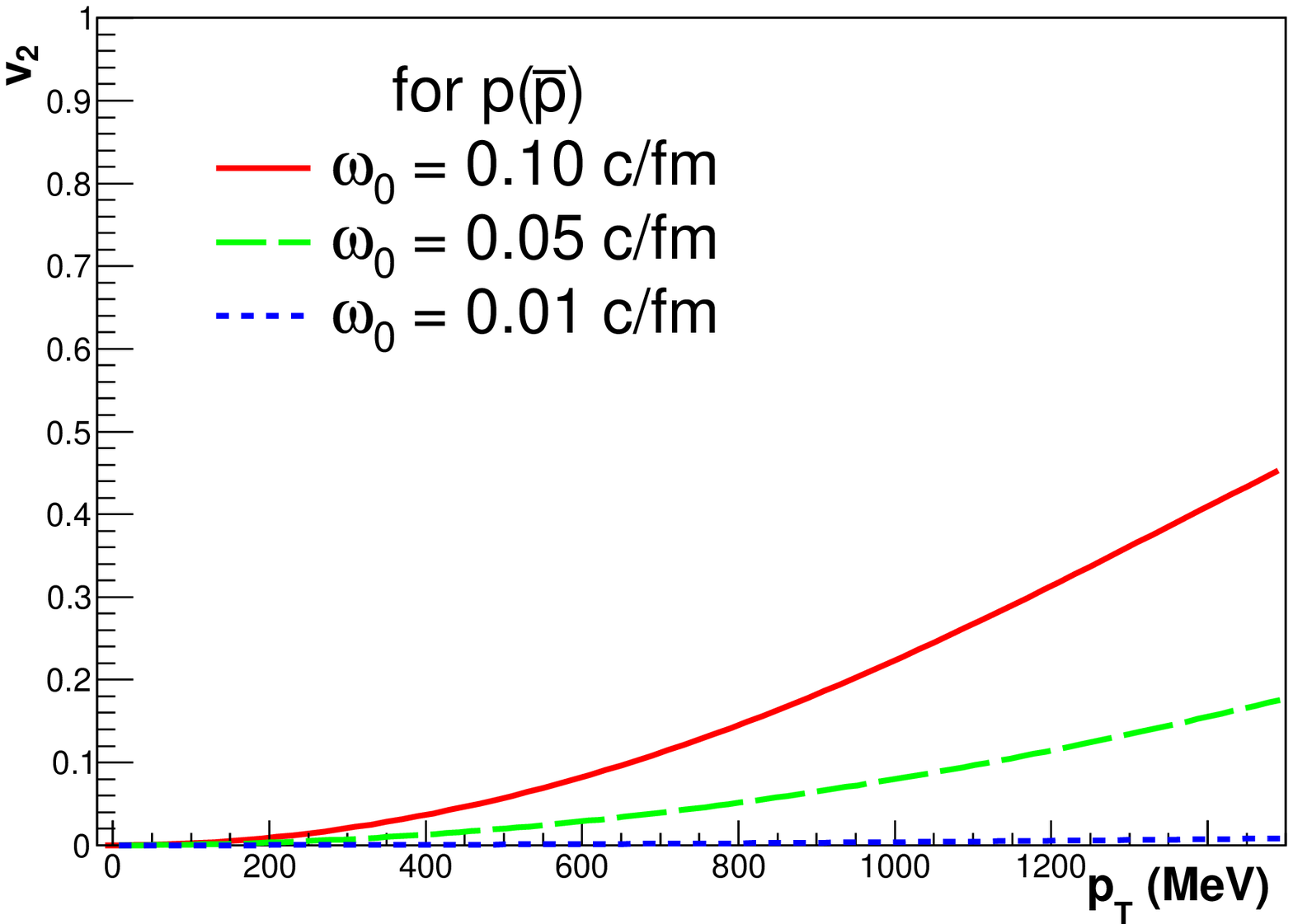}\\
\caption{
$v_2$ for pions (upper panel) and protons (lower panel) as a function of $p_T$
taken at the freeze-out time (when $T$ reaches $T_f = 140$ MeV), for three
different $\omega_0$ values.  Initial conditions are the same as in
Fig.~\ref{f:R}.  The effect of rotation is clearly the increase of the elliptic
flow.
}%
\label{f:fv2}%
\end{center}
\end{figure}

\begin{figure}%
\begin{center}
\includegraphics[width=0.75\columnwidth]{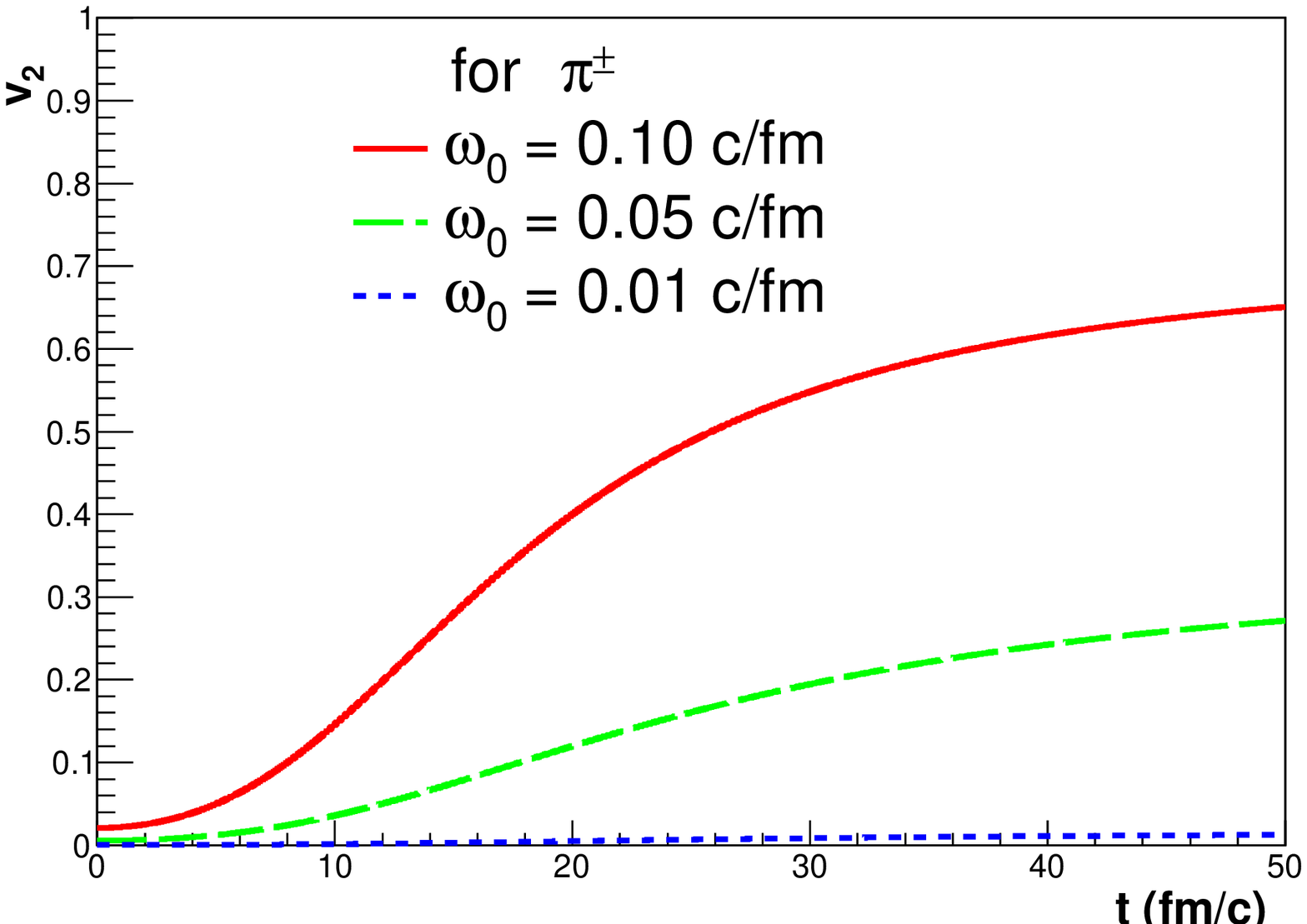}\\
\includegraphics[width=0.75\columnwidth]{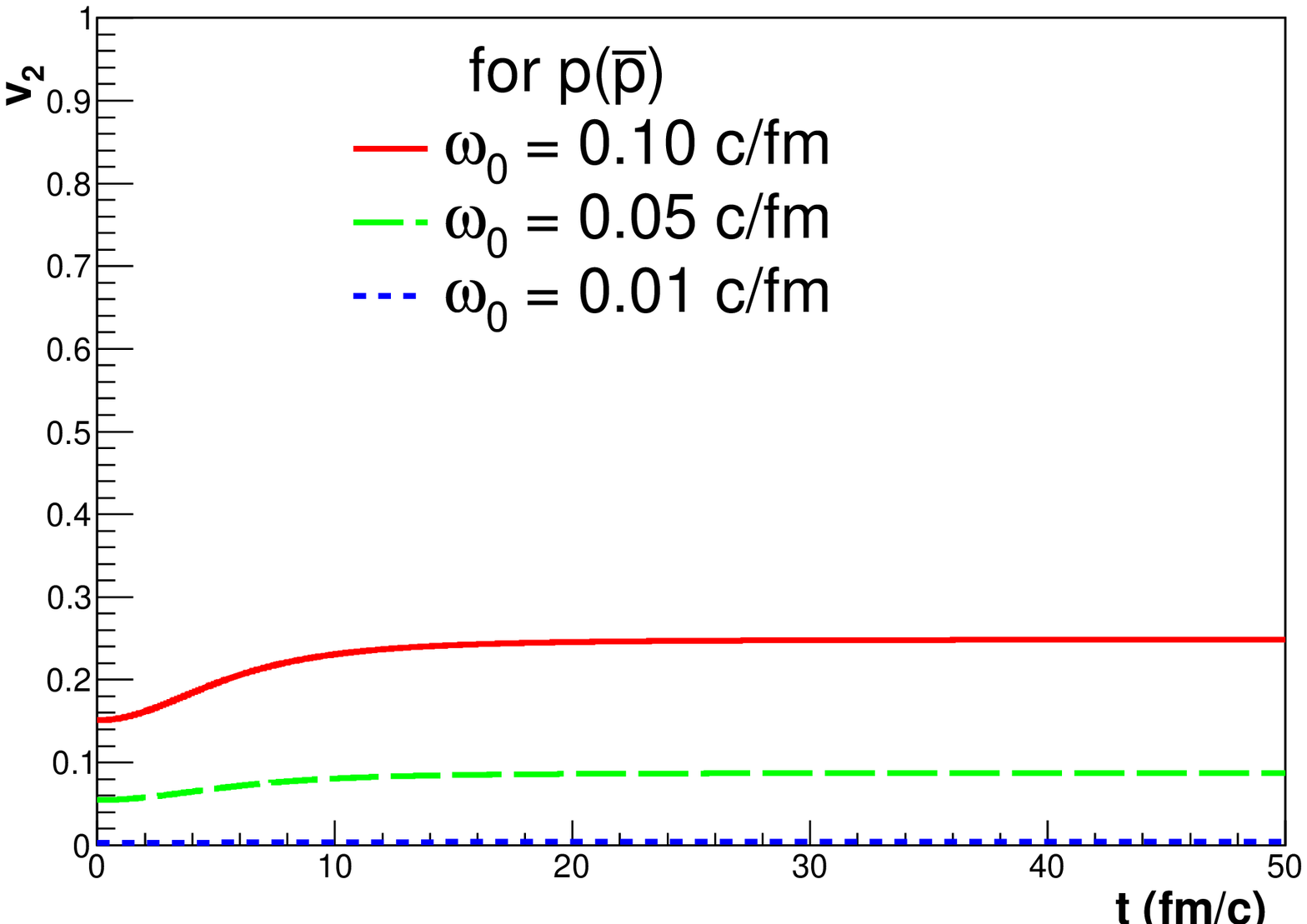}\\
\caption{
Freeze-out time dependence of $v_2$ for pions (upper panel) and protons (lower
panel), at a fixed $p_T$ (300 MeV/c for pions and 1000 MeV/c for protons), for
three different $\omega_0$ values.  Initial conditions are the same as in
Fig.~\ref{f:R}.  The effect of rotation is again seen to increase the elliptic
flow.
}%
\label{f:grv2}%
\end{center}
\end{figure}

Figs.~\ref{f:fv2} and \ref{f:grv2} illustrate again the case of the $X_0 = Y_0
= Z_0 = 5$ fm, $\dot X_0 = \dot Y_0 = \dot Z_0 = 0$, $T_0 = 350$ MeV initial
conditions, with three different values of initial angular velocity $\omega_0$,
to see how rotation influenced the evolution of the elliptic flow.  In this
particular case, we emphasize that the initial geometry is spherical, so
elliptic flow develops as a dynamical effect entirely due to the angular
momentum of an intially spherical volume. In a more realistic situation, where
also initial geometrical asymmetries may be present, one thus expects that the
elliptic flow also will be quite significantly influenced by the initial
angular momentum of the participant zone.

Fig.~\ref{f:fv2} shows the values of $v_2$ for pions and protons (to display
the effect of particle mass) taken at the freeze-out time, as a function of
$p_T$, while Fig.~\ref{f:grv2} shows the $v_2$ of pions and protons for a
representative $p_T$ value (300 MeV/c for pions, 1000 MeV/c for protons), as a
function of freeze-out time. It is clear that in this demonstrative case of
spherically symmetric initial conditions, rotation is what gives $v_2$ its
magnitude. It is also straightforward then to conclude that the non-vanishing 
initial angular momentum of the participant zone may be an
important contributor to the experimentally found and rather  
significant elliptic flow $v_2$ of observed particles.

%
%

\subsection{Two-particle correlations}

The two-particle Bose-Einstein correlation function (BECF) can be calculated
from the Fourier-transform of the source function $S\z{t,\v{r},\v{p}}$ of
Eq.~(\ref{e:Sdef}).  For an introduction and review on this topic and the
evaluation of Bose-Einstein correlation functions from analytic and realistic
hydrodynamical models, we recommend~\cite{Csorgo:1999sj}.  The simplest
(spherically symmetric, nonrelativistic, non-rotational) examples for this kind
of calculations are given in Refs.~\cite{Csorgo:1994fg,Csizmadia:1998ef}, that
we reproduce here in the special limit of zero initial angular momentum and
spherical symmetry.  Furthermore, utilizing the core-halo
picture~\cite{Csorgo:1994in} one introduces the $\lambda$ parameter, the
effective intercept of the correlation function: $\lambda\equiv\lambda\z{\v{p}}
= \sz{N_c\z{\v{p}}/N\z{\v{p}}}^2$.  The $\lambda$ parameter measures the
fraction of particles originating directly from the core, in contrast to those
stemming from decay of longer lived resonances. The two-particle correlation
function turns out to take the form

\begin{eqnarray}
 C\z{\v{K},\v{q}} & = & 
	1 + \lambda \exp\kz{- q_x^2 R_x^2 - q_y^2 R_y^2 - q_z^2 R_z^2} , \label{e:ellbecf}\\
 \v{K} & = & 
	\v{K}_{12}\, = \, \frac{1}{2}\z{\v{p}_1 + \v{p}_2} , \\
 \v{q} & = & 
	\v{q}_{12}\, = \, \z{\v{k}_1 - \v{k}_2} , \\
 R_x^{2} & = & 
	X_f^{2} \z{ 1 + \frac{m}{T_f} \z{\dot X_f^2 { + Z_f^2 \omega_f^2}}}^{-1} 
	\, = \,
	R_f^{2} \z{ 1 + \frac{m}{T_f} \z{\dot R_f^2 { + R_f^2 \omega_f^2}}}^{-1} 
		, \label{e:rxp}\\
 R_y^{2} & = & 
	Y_f^{2} \z{ 1 + \frac{m}{T_f}    \dot Y_f^2 }^{-1} , \label{e:ryp}\\
 R_z^{2} & = & 
	Z_f^{2} \z{ 1 + \frac{m}{T_f} \z{\dot Z_f^2 { + X_f^2 \omega_f^2}} }^{-1} 
	\, = \,
	R_f^{2} \z{ 1 + \frac{m}{T_f} \z{\dot R_f^2 { + R_f^2 \omega_f^2}}}^{-1} 
		.  \label{e:rzp}
\end{eqnarray}

Remember again, that although we have written these radius parameters in a
symmetric way, the spheroidal symmetry of the applied solution, $X = Z$, $\dot
X = \dot Z$, actually impies $R_x = R_z$. These radius parameters can also be
called the lengths of homogeneity~\cite{Sinyukov:1994vg}.
Again, note the very simple occurrence of rotational terms 
in the expression of the radius parameters, just
as it was for the slope parameters $T_x=T_z$ and $T_y$ in Eqs.~(\ref{e:txp})
and (\ref{e:tzp}). These terms imply that increased initial angular momentum
leads to an enhanced mass dependence and a decrease in the HBT radius parameters.
This analytic result yields a deep insight into the earlier, 
numerically obtained indication that an increased rotation leads to 
a decreased HBT radius parameter, Ref.~\cite{Velle:2015dpa}.

Up to now we have dealt only with the case when the particle emission happens
instantaneously (the delta function in \Eq{e:Sdef} assures this). For
Bose-Einstein correlations, it is instructive to see what happens when the
freeze-out has a small but finite duration time. Let this duration be denoted
by $\Delta t$. We evaluate the correlation functions by setting the time
dependence of the source function equal to $(2\pi\Delta
t^2)^{-1/2}\exp[-(t-t_f)^2/ 2\Delta t^2]$, instead of $\delta(t-t_f)$.
Performing the calculation again, one arrives at the conclusion that all the
previous radius components, including the cross-terms, are supplemented with an
additional term $\delta R^2_{ij} = \beta_i\beta_j \Delta t^2$, where
$\mbox{\boldmath$\beta$}=({{\bf p}_1+{\bf p}_2})/ ({E_1 + E_2})$ is the
velocity of the pair. 

We write up the Bose-Einstein correlation functions in a more usual
parametrization, the so-called side-out-longitudinal or Bertsch-Pratt (BP)
parameterization. In this parametrization, the longitudinal direction, $r_{\rm
long}\equiv r_{\rm l}$ coincides with the beam direction, the . The ,,out''
direction, $r_{{\rm out}}\equiv r_{\rm o}\,$ is parallel to the mean transverse
momentum of the pair in the longitudinal co-moving frame (LCMS), and the side
direction, $r_{\rm side} = r_{\rm s}\,$ is orthogonal to both of these. The
mean velocity of the particle pair can be written in the Bertsch-Pratt system
as $\mbox{\boldmath $\beta$}= (\beta_{\rm o},0,\beta_{\rm l})$, where
$\beta_{\rm o} = \beta_t\,$. We denote by $\phi$ the angle of the event plane
and the mean transverse momentum of the measured pair. Performing the
coordinate transformation we obtain the following result:

\begin{eqnarray}
 C_{2}\z{\v{K},\v{q}} &=& 1+\lambda \exp \z{-\sum_{i,j={\rm s,o,l}}q_i q_j R_{ij}^2},\\
 R^2_{\rm s}  &=& R_y^2 \cos^2\phi + R_x^2 \sin^2\phi\,, \label{e:R2s}\\
 R^2_{\rm o}  &=& R_x^2 \cos^2\phi + R_y^2\sin^{2}\phi + \beta_{t}^{2}\Delta t^{2}, \\
 R^2_{\rm l}  &=& R_x^2 + \beta _{\rm l}^{2}\Delta t^{2}, \\
 R^2_{\rm ol} &=& \beta _t\beta _{\rm l}\Delta t^{2}, \\
 R^2_{\rm os} &=& \z{R_x^2-R_y^2} \cos \phi \sin \phi\,, \\
 R^2_{\rm sl} &=& { 0} .\label{e:R2sl}
\end{eqnarray}

These results are simple and straightforward generalizations of the
spheroidally symmetric special case of Ref.~\cite{Csorgo:2001xm}.  As a
function of $\phi$, many of these radius parameters oscillate. Again, it is
important to emphasize that the fact that eg. $R_{\rm sl} = 0$, and that
$R_{\rm ol} = 0$ for $\Delta t = 0$, is not some general result, but rather an
artefact of our considered class of spheroidal solutions where $X = Z$. In a
more general case of $X \neq Z$, these cross-terms would appear (as seen eg.
from the results for a non-rotating but tilted ellipsoid in
Ref.~\cite{Csorgo:2001xm}.) The effect of rotation, however, enters in a very
simple way: it results in the decrease of the radius parameters compared to the
non-rotating case, as follows analytically from Eqs. ~(\ref{e:rxp}-\ref{e:rzp})
that are related the observable HBT radii through Eqs.~(\ref{e:R2s}-\ref{e:R2sl}).

It is necessary to note that in the non-relativistic model investigated in this
work, we see that the HBT radius parameters do not depend on the transverse
momentum of the particle pair at all. To properly account for the observed
$m_T$ dependence of the radius parameters, one needs to invoke relativistic
dynamics; for example, in the Buda-Lund parametrizations of Refs.
~\cite{Csanad:2003qa},\cite{Csorgo:1995bi} it
turns out that the functional form of the dependence on particle mass $m$ seen
in Eqs.~(\ref{e:R2s})--(\ref{e:R2sl}) is approximately conserved, just in the
relativistic case, the dependence is on the transverse mass $m_T$ of the
particle pair. These parameterizations evolved to various hydrodynamical
solutions, in particular the present manuscript can be considered
as a non-relativistic, rotating and spheroidally symmetric Buda-Lund hydrodynamical solution. So, similarly to the remarks after \Eq{e:Teffdef}, we conclude
that as long as no relativistic description of the observables in a rotating
system is available, our results give a hint at the proper dependence of the
radius parameters on the kinematical variables by setting $m\to m_T$ in the
formulas above.

\begin{figure} 
\begin{center}
\includegraphics[width=0.50\columnwidth]{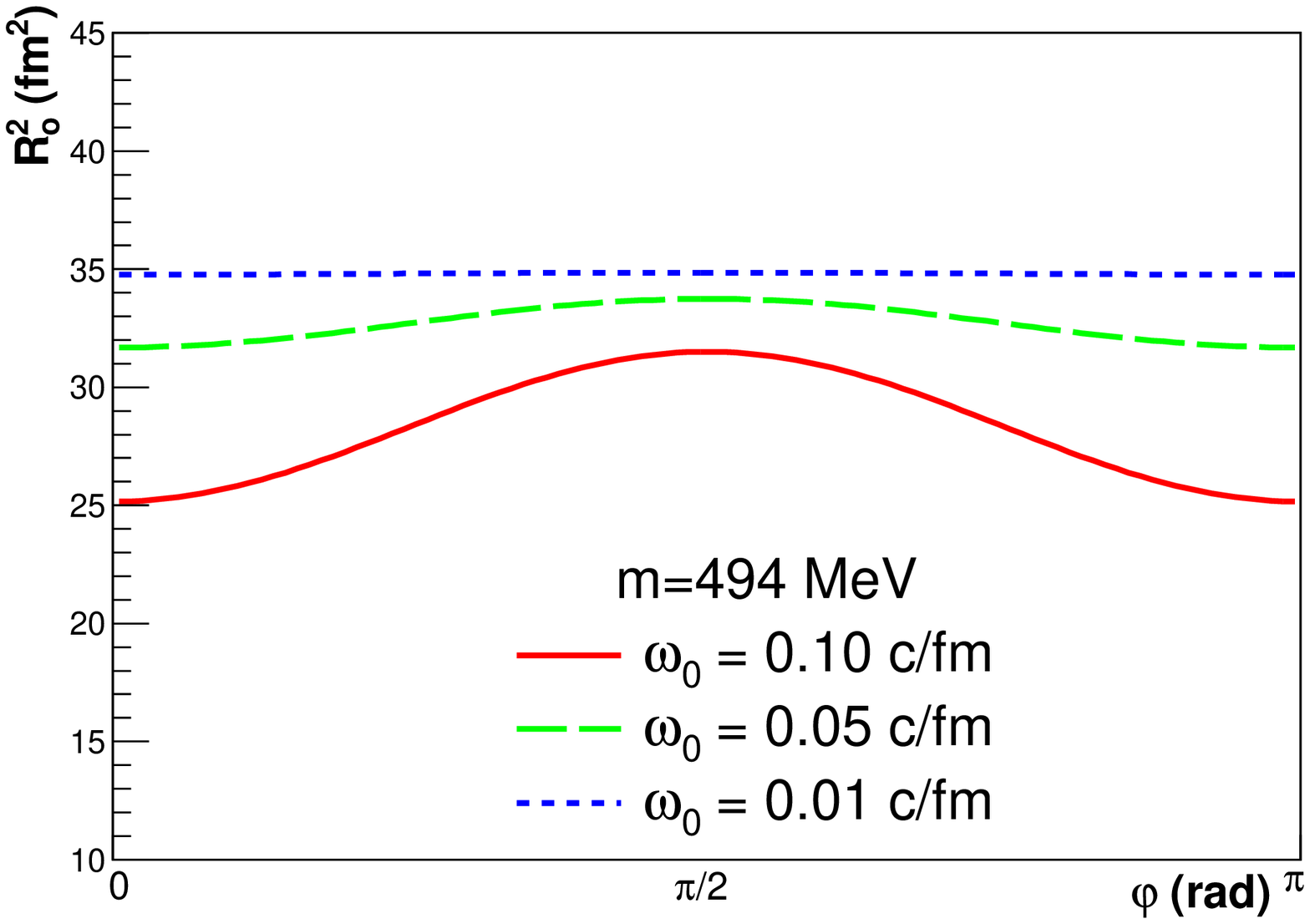}\\
\includegraphics[width=0.50\columnwidth]{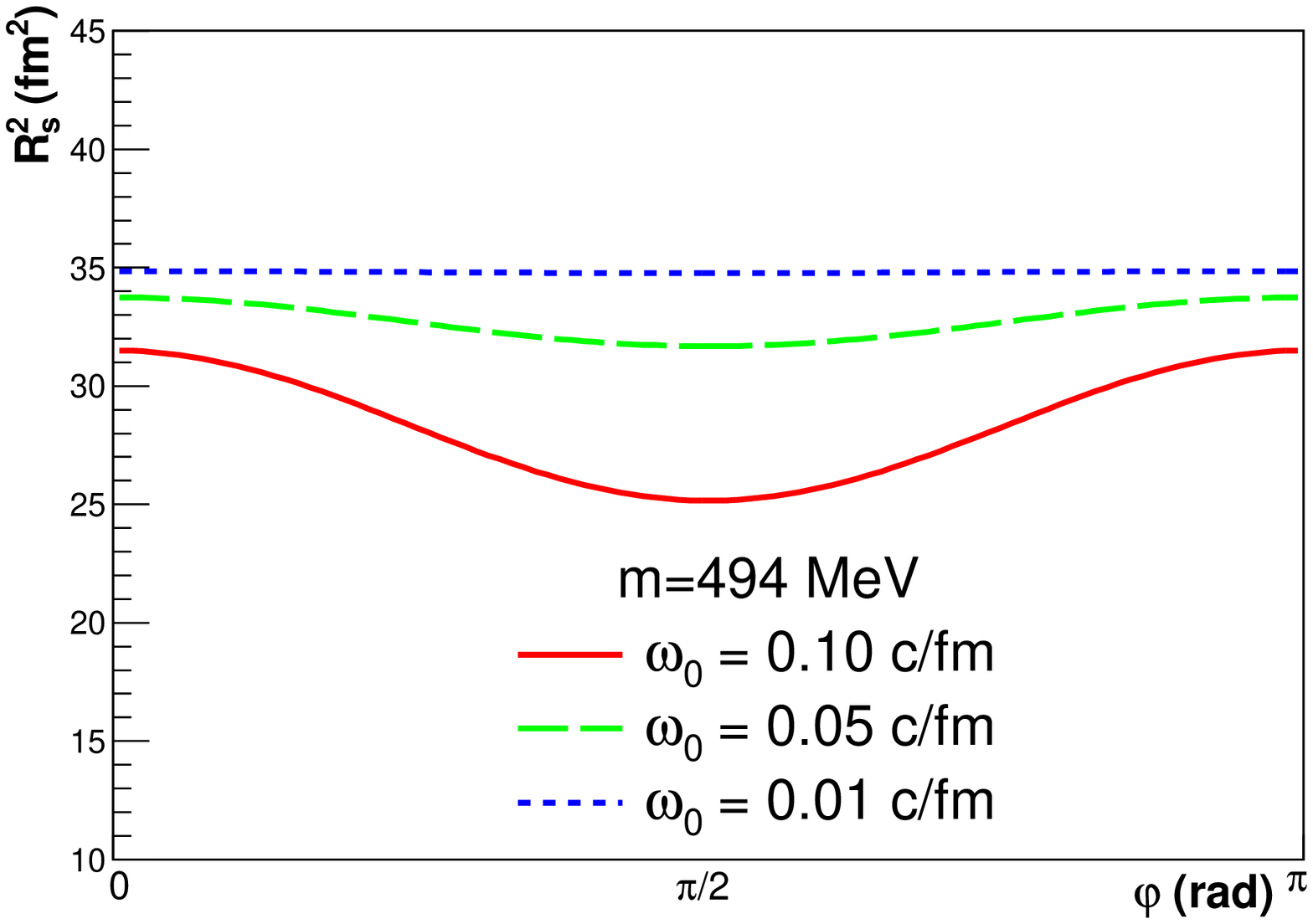}\\
\includegraphics[width=0.50\columnwidth]{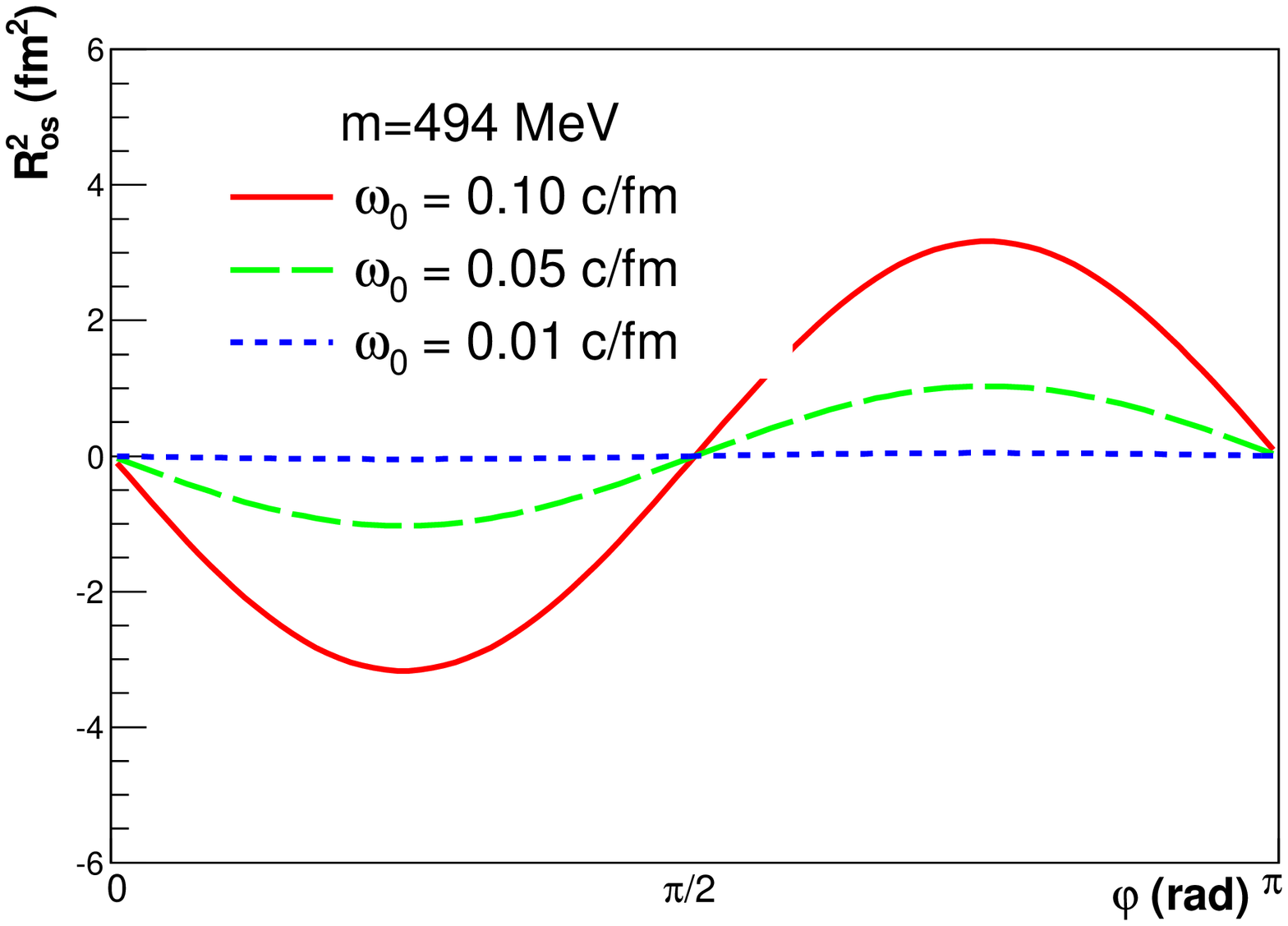}\\ 
\caption{ 
HBT radius parameters $R^2_{\rm o}$ (upper panel), $R^2_{\rm s}$ (middle panel)
and the cross-term $R^2_{\rm os}$ (lower panel) as a function of the azimuthal
angle of the pair, at freeze-out.  Initial conditions are the same as in
Fig.~\ref{f:R}, but this  observable was calculated for charged kaons, with
mass $m = 494$ MeV.  The cases with three different $\omega_0$ values
illustrate the effect of rotation: it causes more anisotropy as well as an
overall decrease of the magnitude of the azimuthally averaged radii.
}%
\label{f:fR2}%
\end{center}
\end{figure}

\begin{figure}%
\begin{center}
\includegraphics[width=0.75\columnwidth]{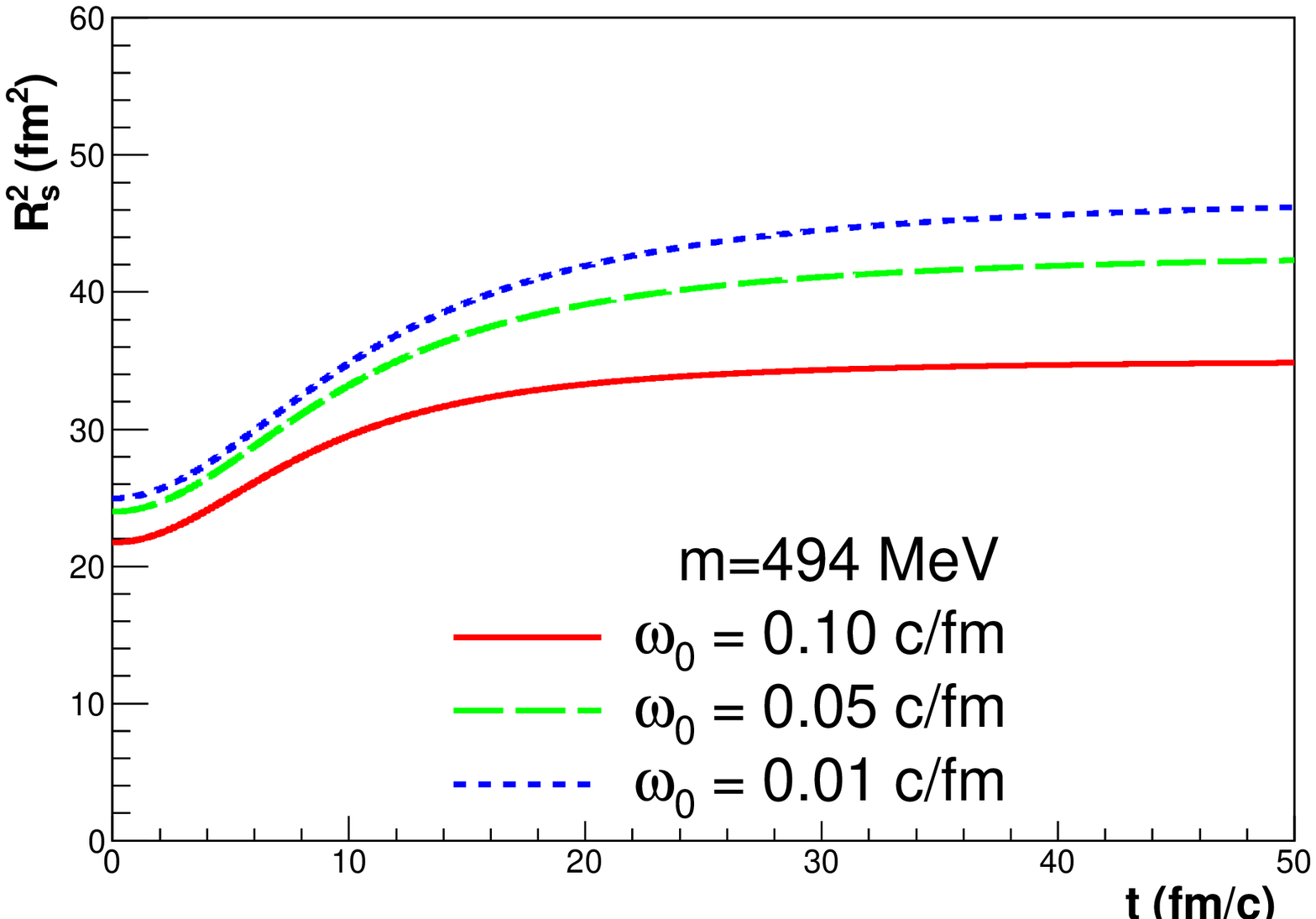}\\
\includegraphics[width=0.75\columnwidth]{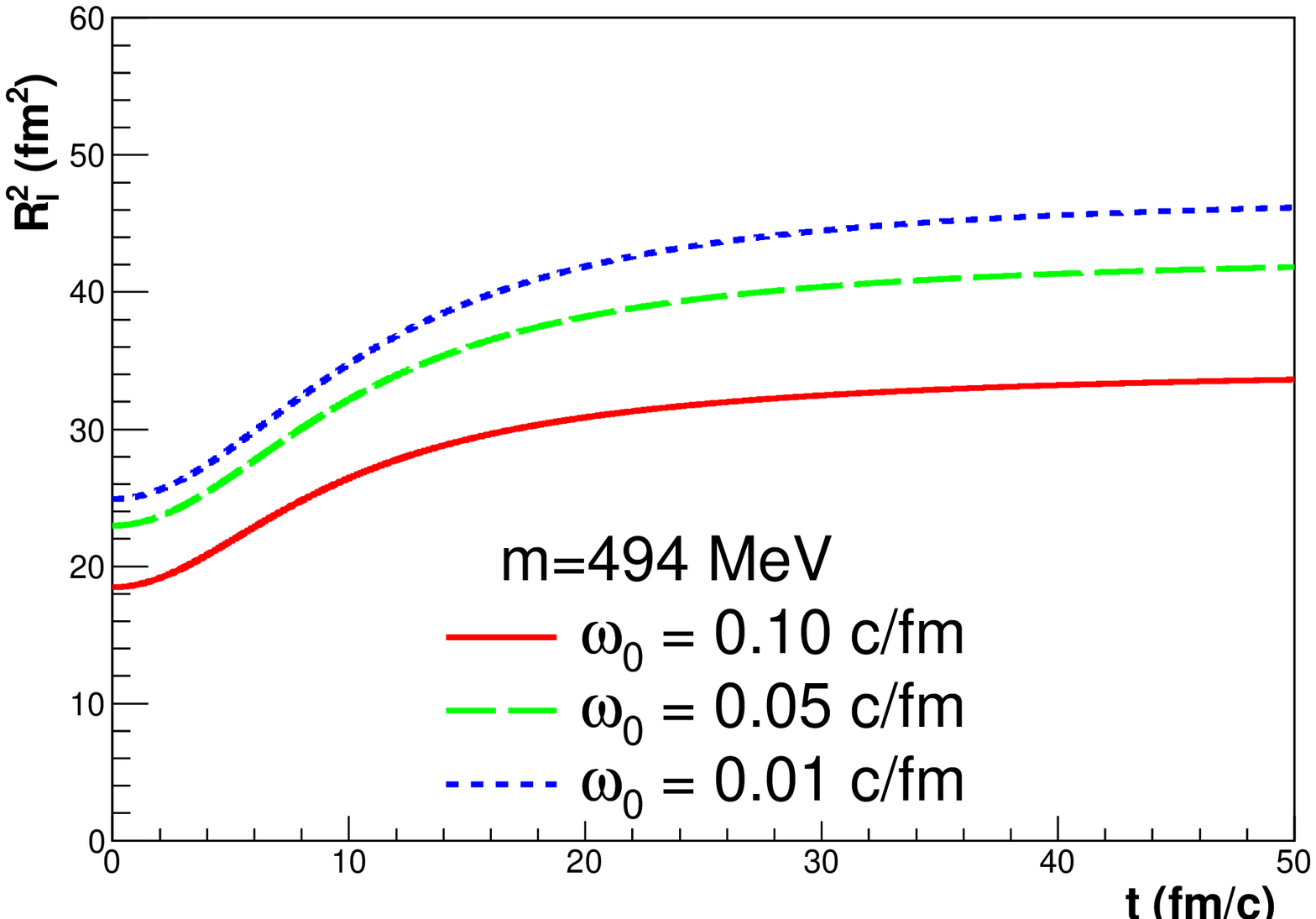}\\
\caption{
Freeze-out time dependence of the azimuthal average of $R^2_{\rm s}$ (upper
panel) and $R^2_{\rm l}$ (lower panel).  Initial conditions are the same as in
Fig.~\ref{f:R}, and three different $\omega_0$ values.  but this  observable
was calculated for charged kaons, with mass $m = 494$ MeV.  Rotation decreases
the magnitude of the radii.
}%
\label{f:grR2}%
\end{center}
\end{figure}

As before, we illustrate the effects of rotation on the HBT radius parameters
by taking a spherically symmetric initial condition with three different values
of initial angular velocity. Fig.~\ref{f:fR2} shows the azimuthal angle
dependence of $R^2_{\rm o}$, $R^2_{\rm s}$, and $R^2_{\rm os}$ at the
freeze-out time, while Fig.~\ref{f:grR2} shows the dependence of the azimuthal
mean value of $R^2_{\rm o}$ and $R^2_{\rm l}$ on the freeze-out time. (We plot
only these ones, as the others are either not interesting or just very similar
to these.) It is clearly seen that the overall magnitude of the azimuthal mean
of the radius parameters decrease with rotation, as mentioned above. One also
observes that all the oscillations become stronger with more rotation, a
straightforward effect of anisotropic geometry stemming from the centrifugal
force. This behavior is thus expected to be a general consequence of rotation.

\section{Summary}

We have evaluated the single-particle spectra, the elliptic and higher order flows,
and the two-particle Bose-Einstein correlation functions for 
a rotating and expanding, spheroidally symmetric fireball, a non-relativistic
Buda-Lund type hydrodynamical model. 

A non-vanishing  value of the initial angular momentum, 
an important conserved quantity that is characteristic
to the non-central heavy ion collisions, was taken into account and
its effects on the observables have been found using simple and
straight-forward analytic formulae.

Although the hydrodynamical solution we used is
non-relativistic, the insight it gives the observables in rotating systems is
valuable and relevant for analyzing experimental measurements of the treated
observables.

We have proven that rotational terms appear in the observables in a way that is
rather similar to radial flow effects and lead to the increased mass dependence
of the effective temperatures or slope parameters of the single particle
spectra, an enhanced elliptic flow and a stronger mass-dependence of the
decrease of the Bose-Einstein correlation radii. We have illustrated these
analytic results with several plots utilizing an academic, but rather interesting
spherical initial condition and initial angular velocities gradually increasing
from a nearly vanishing value to a realistic value.

Using this special initial condition of a rotating and initially spherical
fireball, have demonstrated that initial angular momentum leads to a
significant elliptic flow even for the case when there are no initial spatial
asymmetries present. This implies that future experimental data analysis has to
take possible rotational effects into account as they may mix in a subtle and
difficult to identify manner with other, more generally considered radial flow
effects.

We also emphasize that to consider these important angular momentum and
rotational effects, finite 3d hydrodynamical solutions have to be utilized for
the data analysis, as infinite systems have infinite moments of intertia so
they cannot realistically model rotating hydrodynamical solutions.

\section*{Acknowledgements}
We greatfully acknowledge inspiring discussions with M.\ Csan\'ad,  L.\ Csernai
and Y. Hatta.  The research of M. N. has been supported by the European Union
and the State of Hungary, co-financed by the European Social Fund in the
framework of T\'AMOP 4.2.4.  A/1-11-1-2012-0001 ''National Excellence" Program.
This work has been supported by the OTKA NK 101438 grant of the Hungarian
National Science Fund. 

\appendix

\section{Equations of motion for $\kappa = 3/2$}\label{s:app:H}

For $\kappa = 3/2$, the solution of the equations of motion, \Eq{e:RYtime} can
be given using the Hamilton-Jacobi formalism, applied to the \r{e:Hamilton}
Hamiltonian. This formalism can be applied for a classical Hamiltonian
$H\z{P_i,Q_i,t}$ as follows. (Here $P_i$ and $Q_i$ denote the $N$ canonical
momenta and coordinates.) One must find the classical Hamilton function, or
action $S\z{Q_i,t}$ that is a solution of the Hamilton-Jacobi equation:
\[
\frac{\partial S}{\partial t} + H\z{\frac{\partial S}{\partial Q_i}, Q_i,t} = 0 .
\]

If a suitable solution $S\z{Q_i,t,K_i}$ is found, which has $N$ different
arbitary parameters $K_i$, then the equations $\frac{\partial}{\partial
K_i}S\z{Q_i,t,K_i} = L_i$, again with arbitary constants $L_i$, give the
solution of the equation of motion.

In cases where $H$ is time-dependent, the conserved energy $E$ enters as a free
parameter:
\[
S\z{Q_i,t} = -Et + S\z{Q_i},\quad H\z{\frac{\partial S(Q_i)}{\partial Q_i}, Q_i} = E.
\]

In our case, for $\kappa = 3/2$, we have
\[
\frac{\z{\frac{\partial S}{\partial R}}^2 + 2\z{\frac{\partial S}{\partial Y}}^2}{4m} +
\frac{m\omega_0^2R_0^4}{R^2} + \frac{3T_0}{2}\z{\frac{R_0^2Y_0}{R^2Y}}^\frac{2}{3} = E .
\]

This can be solved by introducing the $\rho$, $\chi$ variables (like planar
cylindrical coordinates) as 
\[
R = \frac{\rho}{\sqrt{2}}\cos\chi, \quad Y = \rho\sin\chi ,
\]
and re-arranging the Hamilton-Jacobi equation to obtain
\[
\z{\frac{\partial S}{\partial\rho}}^2 + \frac{1}{\rho^2}\kz{\z{\frac{\partial S}{\partial\chi}}^2 + F\z{\chi}} = 2 m E ,
\]
with
\[
F\z{\chi} = 2 m^2\omega_0^2R_0^4 + 3 m T_0\z{\frac{2R_0^2Y_0}{\cos^2\chi\sin\chi}}^{2/3} .
\]

The desired solution is now obtained straightforwardly, by introducing the $K$
constant so that the equation becomes separable.  We thus can assume that
$S\z{\rho,\chi} = S_1\z{\rho} + S_2\z{\chi}$, and get the following solution:
\[
\z{\frac{\partial S_1}{\partial\rho}}^2 + \frac{K}{\rho^2} = 2 m E ,\quad 
\z{\frac{\partial S_2}{\partial\chi}}^2 + F\z{\chi} = K \quad\Rightarrow
\]
\[
S = -Et + \int\m{d}\rho\sqrt{2 m E-\frac{K}{\rho^2}} + \int\m{d}\chi\sqrt{K-F\z{\chi}} .
\]
The equations of motion now can be considered as solved. Taking the
$\frac{\partial S}{\partial E} = B_1 = $const condition, with elementary
integration we get the following result:
\begin{equation}
\rho^2 = \frac{2E}{m}\z{t+B_1}^2 + \frac{K}{2 m E} .
\label{e:rhosol}
\end{equation}

If one writes back $R$ and $Y$ in this result, and chooses the $K$ and $B_1$
constants to match the initial conditions, one gets the following:
\begin{equation}
2R^2+Y^2 = \frac{2E}{m}t^2 + \z{4\dot R_0R_0+2\dot Y_0Y_0}t + 2R_0^2+Y_0^2 ,
\label{e:1stintegral}
\end{equation}
where the $E$ energy is in this case
\[
\frac{E}{m} = \dot R_0^2+\frac{\dot Y_0^2}{2}+\omega_0^2R_0^2+\frac{3T_0}{2m} .
\]
Substituting $E$ back in \Eq{e:1stintegral}, one readily arrives at
\Eq{e:R2Y2}. This gives the time dependence of $2R^2+Y^2$.  This result could
have been inferred by simple inspection of the differential equations, and once
this is known, a univariate differential equation also can be given (and
numerically solved) for the remaining independent variables. This was known
already in Ref~\cite{Akkelin:2000ex} (although not for a rotating system). Our
method here gives the solution by quadratures: it is easy to see that the other
condition coming from the Hamilton-Jacobi method, $\frac{\partial S}{\partial
K} = B_2 = $const, does not contain the time variable $t$. So this condition
fixes the shape of the trajectory in the $R$--$Y$ ,,plane''. However, in
general it cannot be evaluated analytically.

\section{A survey of analytic solutions}\label{s:app:survey}

In this Appendix we give a short list of some recent exact analytic solutions of the non-relativistic hydrodynamical equations.
We do this in the hope that just as uniting efforts in heavy-ion physics and analytical hydrodynamics has been proven to be of
much interest for both research directions, it can and will prove fruitful also in future applications (even though the
heavy-ion physical applications of the mentioned solutions is not clear yet).

In spite of the lacking basic mathematical theorems (like existence and uniqueness) there is a fairly great amount of analytic solutions available
for various hydrodynamical equations like the  Euler or the Navier-Stokes(NS) equations. 
high-energy physics we give a short overview about the recent developments in analytical hydrodynamics.

There are various examination techniques available with numerous studies in the literature.
Most of them are based on the application of the self-similar Ansatz or on Lie algebra methods.
 
Sedov in his classical work about self-similar solutions \cite{sedov} presents one of the first analytic solutions for the
three dimensional spherical NS equation where all three velocity components and the pressure have polar angle dependence
($\theta$) only.
Some similarity reduction solutions of the two dimensional incompressible NS equation was presented by Xia-Yu \cite{jiao}.
Additional solutions are available for the 2+1 dimensional NS also via symmetry reduction techniques by \cite{fakhar}.
Manwai  \cite{manwai} studied the N-dimensional $(N \ge 1)$ radial Navier-Stokes equation with different kind of viscosity and pressure 
dependences and presented analytical blow up solutions. His works are still 1+1 dimensional (one spatial and one time dimension) investigations.

In our former study we generalized the self-similar Ansatz of Sedov and presented analytic solutions for the most general 
three dimensional Navier-Stokes equation in Cartesian coordinates \cite{imre}. The solutions are the Kummer functions with 
quadratic arguments. Later, we applied our method to the three dimensional compressible NS equation where the politropic
equation of state was used \cite{imre2} resulting in the Whitakker functions. 
 
Recently, Hu {\it{et al.}} \cite{hu} presented a study where symmetry reductions and exact solutions of the (2+1)-dimensional NS were presented. 
Aristov and Polyanin \cite{arist} use various methods like Crocco transformation, generalized separation of variables or the method of
functional separation of variables for the NS and present large number of new classes of exact solutions. 

A universal Lie algebra study is available for the most general three dimensional NS \cite{lie}. Unfortunately, no explicit solutions
are shown and analyzed there. Fushchich {\it{et al.}} \cite{fus} constructed a complete set of ${\tilde{G}}(1,3)$-inequivalent
Ans\"atze of codimension one for the NS system, they present 19 different analytical solutions for one or two space dimensions.  
Further, two and three dimensional studies based on group theoretical method were presented by Grassi \cite{grassi} getting 
Whitakker functions for solutions which are very similar to our results mentioned earlier \cite{imre}.

\end{document}